\documentclass[longauth]{aa}  
\usepackage[colorlinks=true]{hyperref}
\hypersetup{linkcolor=blue,citecolor=blue,filecolor=cyan,urlcolor=blue}

\usepackage{graphicx}
\usepackage{txfonts}
\begin{document}



\titlerunning{FAUST. XXVIII. High-Resolution ALMA Continuum Observations of Class 0/I Disks}
   \title{FAUST. XXVIII. High-Resolution ALMA Observations of Class 0/I Disks: Structure, Optical Depths, and Temperatures}
   \authorrunning{Maureira et al.}
   \author{M. J. Maureira\inst{1}, J. E. Pineda \inst{1}, H. B. Liu \inst{2}, P. Caselli \inst{1}, C. Chandler \inst{3}, L. Testi \inst{4,5},  D. Johnstone \inst{6,7}, D. Segura-Cox \inst{8,1}, L. Loinard \inst{9,10,21}, E. Bianchi \inst{5},C. Codella \inst{5,11}, A. Miotello \inst{12},L. Podio \inst{5}, L. Cacciapuoti \inst{13}, Y. Oya \inst{14}, A. Lopez-Sepulcre \inst{11,15}, N. Sakai \inst{16}, Z. Zhang \inst{16}, N. Cuello \inst{11}, S. Ohashi \inst{17},  Y. Aikawa \inst{18}, G. Sabatini \inst{5} , Y. Zhang \inst{19}, C. Ceccarelli \inst{11},
          \and
         S. Yamamoto \inst{20}
          }

    \institute{Max-Planck-Institut für extraterrestrische Physik (MPE), Gießenbachstr. 1, D-85741 Garching, Germany, 
              \email{maureira@mpe.mpg.de}
         \and Department of Physics, National Sun Yat-Sen University, No. 70, Lien-Hai Road, Kaohsiung City, 80424, Taiwan, ROC ; Center of Astronomy and Gravitation, National Taiwan Normal University, Taipei, 116, Taiwan, ROC 
        \and National Radio Astronomy Observatory, PO Box O, Socorro, NM 87801, USA 
         \and Dipartimento di Fisica e Astronomia “Augusto Righi”, Viale Berti Pichat 6/2, Bologna, Italy 
         \and INAF, Osservatorio Astrofisico di Arcetri, Largo E. Fermi 5, 50125, Firenze, Italy 
 \and NRC Herzberg Astronomy and Astrophysics, 5071 West Saanich Road, Victoria, BC V9E 2E7, Canada 
          \and Department of Physics and Astronomy, University of Victoria, Victoria, BC V8P 5C2, Canada 
            \and Department of Physics and Astronomy, University of Rochester, Rochester, NY 14627, USA 
        \and Instituto de Radioastronom\'ia y Astrof\'isica, Universidad Nacional Aut\'onoma de M\'exico, Morelia, 58089, M\'exico 
          \and Black Hole Initiative at Harvard University, 20 Garden Street, Cambridge, MA 02138, USA 
                    \and Univ. Grenoble Alpes, CNRS, IPAG, 38000 Grenoble, France 
           \and European Southern Observatory, Karl-Schwarzschild-Str 2, 85748, Garching, Germany 
           \and European Southern Observatory, Alonso de Cordova 3107, Vitacura,
Region Metropolitana de Santiago, Chile 
            \and Center for Gravitational Physics, Yukawa Institute for Theoretical
Physics, Kyoto University, Oiwake-cho, Kitashirakawa, Sakyo-ku,
Kyoto-shi, Kyoto-fu 606- 8502, Japan 
\and Institut de Radioastronomie Millimétrique, 38406 Saint-Martin
d’Héres, France 
          \and RIKEN Cluster for Pioneering Research, 2-1, Hirosawa, Wako-shi, Saitama 351-0198, Japan 
\and National Astronomical Observatory of Japan, Osawa 2-21-1, Mitakashi, Tokyo 181-8588, Japan 
           \and Department of Astronomy, The University of Tokyo, 7-3-1 Hongo, Bunkyo-ku, Tokyo 113-0033, Japan 
          \and Department of Astronomy, Shanghai Jiao Tong University, 800
Dongchuan Road, Minhang, Shanghai 200240, People’s Republic
of China 
          \and SOKENDAI (The Graduate University for Advanced Studies), Shonan Village, Hayama, Kanagawa 240-0193, Japan 
          \and David Rockefeller Center for Latin American Studies, Harvard University, 1730 Cambridge Street, Cambridge, MA 02138, USA 
             }

   \date{Received June 25, 2025; accepted October 21, 2025}


  \abstract{Measuring the properties of disks around Class 0/I protostars is crucial for understanding protostellar assembly and early planet formation. We present high-resolution ($\sim$7.5 au) ALMA continuum observations at 1.3 and 3 mm of 16 disks around Class 0/I protostars across multiple star-forming regions (Taurus, Ophiuchus and Corona Australis) and variety of multiplicities. Our observations show a wide range of deconvolved disk sizes ($\sim$2–100 au) and the presence of circumbinary disks (CBDs) in all binaries with separations $<100$ au. The measured properties show similarities to Class II disks, including (a) low $\alpha_\text{1.3-3mm}$ values ($\alpha_{\text{disks}}=2.1^{+0.5}_{-0.3}$) that increase with disk radius, (b) 3 mm disk sizes only marginally smaller than at 1.3 mm (<10\%), and (c) radial intensity morphologies well described by modified self-similar profiles. However, there are key differences: (i) $\alpha_\text{1.3-3mm}$ values increasing monotonically with radius, but exceeding 2 only at the disk edge (2) higher brightness temperatures $T_b$, comparable to or higher than the predicted midplane temperatures due to irradiation, and (iii) $\sim10\times$ higher luminosity at a given size compared to the Class II disks. Together, the results confirm significant optical depth in the observed Class 0/I disks, most with $T_\text{bol}<200$ K, at both 1.3 and 3 mm. Assuming fully optically-thick disks at these wavelengths can explain the higher luminosities compared with Class II disks, but the most compact ($\lesssim$ 40 au) disks require also higher temperatures, suggesting additional heating from viscous accretion. Taking into account the high optical depths, most disk dust masses are estimated in the range 30-900 M$_{\oplus}$ (or 0.01-0.3 M$_{\odot}$ in gas), with some disks potentially reaching marginal gravitational instability. Based on the elevated $T^{1.3mm}_b$, the median location of the water iceline is $\sim$3 au, but this location can extend to more than 10–20 au for the hottest disks in the sample. CBDs exhibit lower optical depths at both wavelengths and hence higher spectral index values ($\tau_{3mm}\lesssim1$, $\alpha_{\text{CBD}}=3.0^{+0.2}_{-0.3}$), dust masses of $\sim10{^2}$ M$_{\oplus}$, and dust emissivity indices $\beta_{\text{CBD}}\sim1.5$ (2 Class 0 CBDs) and $\sim$1 (1 Class I CBD), suggesting substantial grain growth only in the more evolved CBD. The high optical depths inferred from our analysis provide a compelling explanation for the apparent scarcity of dust substructures in the younger Class 0/I disks at $\sim1$ mm, despite mounting evidence for early planet formation.

}


\maketitle
%

\section{Introduction}

Growing evidence suggests that the planet formation process begins during the embedded protostellar stages (Class 0/I), making the characterization of protostellar disks key to study both the protostar accretion process and the initial phases of planet formation. The evidence for the early formation of planets comes from high-resolution ($\sim$ 6 au) observations towards the more evolved Class II disks ($\lesssim$1 Myr) that revealed widespread substructures, such as rings and gaps, thought to be linked to planet formation \citep{2015ALMAPartnershipLongBaseline, 2018Andrewsdsharp,2018Longsubstructures}, alongside direct detections of forming planets through continuum emission or gas kinematics \citep{2021BenistyCircumplanetary,2018PinteKinematic,2022IzquierdoNew, 2025Dominguez-JamettMulti}.
Another line of evidence for early planet formation in embedded disks is the very low dust masses measured in more evolved Class II disks around stars and brown dwarfs \citep{2018Manaramass,2016TestiBrownDwarf,2020SanchisDemogra}. These low values are consistent with models of early pebble growth, migration and  loss through radial drift \citep{2023ApelgrenDiscPopulationSynthesis,2025AppelgrenTheEvolution}  which predict a rapid depletion of the millimeter-dust grains in the disk within a few tenths of a Myr, hence also requiring a fast start of the planet formation process. Tentative evidence for a stable or increase in median disk dust masses at around 2–3 Myr is also consistent with early planet formation and subsequent  dust regeneration driven by disk-planet interaction \citep{2019TurriniResurgence,2022TestiPPpopulation,2022BernaboDustResurgence,2025PlychroniResurgence}. At odds with the results of \cite{2025AppelgrenTheEvolution}, recent multiwavelength studies have challenged the standard assumption that Class II disks emit optically thin radiation at $\sim$1 mm \citep{2021MaciasTW,2024Guerra-AlvaradoBand9HLTau,2024ChungClassII,2025PainterDenseSED,2025GarufiCentSED}. Accounting for optically thick Class II disks provides a potential alternative scenario that still supports early planetesimal formation, since reproducing the observed fluxes and disk sizes would then require the presence of early ($\lesssim$0.4 Myr) dust traps \citep{2024DelussuPop}.\\



Surveys of Class 0/I protostars have progressively improved in resolution, helping in the identification and characterization of the dust emission at millimeter wavelengths from these young disks (e.g., \citealt{2007JorgensenPROSAC,2016TobinVLA,2018SeguraCoxVandamV,2019MauryRare,2020TobinOrion,2023OhashieDisk,2024HsiehCAMPOSI, 2024ReynoldsPerseus}). Recently, more focus has been placed on determining when dust substructures, similar to those in Class II disks, first appear \citep{2020SeguraCoxFour,2020NakataniSubstruc,2023OhashieDisk,2024MaureiraSM1,2024Guerra-AlvaradoIRAS4A1, 2025HsiehCAMPOSII_substructure}. Considering individual studies and surveys with resolutions of $\sim$6–14 au, nearly 60 Class 0/I disks have been observed \citep{2021MCiezaOdiseaIII,2023OhashieDisk,2024HsiehCAMPOSI} and definitive annular substructures have been detected in only five disks (L1489 IRS, ISO-Oph 54, R CrA IRS 2, Oph-emb 20, and IRS 63), all Class I \citep{2018SheehanGY91,2020SeguraCoxFour,2023YamatoL1489IRSedisk,2024ShoshiWL17,2024HsiehCAMPOSI}. \cite{2025HsiehCAMPOSII_substructure} showed that accounting for observational biases, such as inclination and disk size relative to resolution, the detection rate for Class I disks reaches $\sim$60\% at bolometric temperatures of $\sim$250 K, despite the low apparent numbers. Whether optical depth is what further suppresses observed rates, particularly at earlier times, remained unclear. These results suggest either that planet formation begins during the Class I stage or that many younger disks remain too optically thick at $\sim1$ mm, preventing the clear detection of substructures. In fact, \cite{2020NakataniSubstruc} reported that a possibly annular substructure in the Class 0/I disk L1527 is detected 7 mm but not at 1 mm and 3 mm. Similarly, \cite{2024MaureiraSM1} found possible evidence of substructure at 3 mm in a nearly face-on Class 0 disk. Finally, \cite{2021ZamponiHotdisk} and \cite{2024Guerra-AlvaradoIRAS4A1} find possible evidence of substructure in two Class 0 disks that are both massive, hot, and with large scale heights, based on the detection of asymmetries in their profiles. These recent studies, support the notion that wavelength-dependent opacity play a significant role in the observational results. Thus, to build a reliable timeline for planet formation, it is crucial that we quantify the presence and extent of optically thick emission at millimeter wavelenghts in young Class 0/I disks. 

The implications of optically thick emission extend beyond substructure observability. Dust masses, usually derived assuming optically thin and isothermal emission \citep{2007JorgensenPROSAC,2020TobinOrion,2020TychoniecDust}, are then only lower limits.  Optical thickness also biases grain growth estimates, whose sizes may be overestimated if there is significant optically thick emission \citep{2017LiSpectralIndex}. Furthermore, young disks may be heated by shocks and accretion \citep{2015EvansGIDisk,2021ZamponiHotdisk,2021XuFormation,2022Maureirahotspots,2024TakakuwaeDisk}, which can produce very low $\alpha$ values in the presence of optically thick emission, sometimes even below the optically thick limit, without requiring scattering from large-grains \citep{2018GalvanMadridEffects,2021ZamponiHotdisk}.\\

In this study, we present $\sim$7.5 au resolution ALMA continuum observations at 1.3 and 3 mm of Class 0/I disks from the FAUST program\footnote{Fifty AU Study of the chemistry in the disk/envelope system of Solar-like protostars} \citep{2021CodellaFAUST}. The high resolution enabled spatially resolved spectral index maps to probe the extent of the optically thick emission. 
The paper is organized as follows: in Sections~\ref{sec:sample} and~\ref{sec:data} we describe the sample and observations, respectively. In Section~\ref{sec:results}, we present the continuum maps, spectral index maps and corresponding radial profiles. In Section~\ref{sec:analysis}, we quantify the morphology of the intensity profiles at both wavelengths, reveal correlations between disk sizes, disk fluxes and spectral index, and compare dust and brightness temperatures. In Section~\ref{sec:discussion} we discuss how these results directly imply high fractions of optically thick emission at 1.3 and 3 mm. In this context, we discuss and derive implications for the disk masses, temperatures and grain sizes, including circumbinary disk structures. Section~\ref{sec:conclusions} presents the summary and conclusions. 

\begin{table*}
\caption{Detected Class 0 and I protostellar sources properties}
\label{table:source_prop}      
 \centering
\begin{tabular}{l c c c c c c  c c}
\hline\hline
Source & L$_{bol}$ & L$_{int}$& T$_{bol}$ & Class & M${_{prot}}$ & Cloud  & Nearest protostar sep. & FAUST line\\    
 & L$_{\odot}$ &  L$_{\odot}$ & K & & M${_\odot}$ &    &  au &paper\\  
\hline                        %
RCrA IRS7B a,b & 5.4 & 7.4 & 122 & I &2.1-3.2 (a) & RCrA &  $103$ (a-b) &  \cite{2024SabatiniOutflowFAUST} \\   
RCrA SMM1C & 4.7 & $<0.6$ &27 & 0 & - & RCrA &  $803$ (IRS7A)  & "\\ 
RCrA  SMM2 a,b\tablefootmark{a}& 0.6 &0.7& 164 & I & - & RCrA &  $14$ (a-b) & "\\ 
RCrA  IRS7A & 13.9 & 12.9 &125 & I & - & RCrA &  $803$ (SMM1C)  & "\\ 
CrA-24\tablefootmark{a} & 0.1 & - & 370 & I & - & RCrA & $3846$ (SMM1C)& "\\ 
CXO 34& - & - & - & I & - & RCrA &  $1069$  (IRS7B b)& "\\ 
Elias 29& 28.9 & 4.9 & 391 & I & 0.8-1 & Oph &  $>7000$ \tablefoottext{c} & \cite{2025OyaFAUSTElias}\\ 
VLA 1623 A,B & 3.6 & 1.1 & 35 & 0 & 0.3-0.5 (Aa+Ab) & Oph &   $30$ (Aa-Ab) & \cite{2022OhashiVLA1623}\\ 
& & &  &  & 0.4-1.7 (B) &  &  $146$ (Ab-B) & \cite{2024CodelaVLA1623Streamer}\\  
VLA 1623 W & 0.14 & 0.3 & 84 & I & 0.5 & Oph &  1271 (VLA 1623B)  & \cite{2023MercimekVLA1623W}\\ 
Oph A SM1\tablefootmark{a} & < 1 & < 1 & 40\tablefootmark{b} & 0 & - & Oph & 2686 (VLA1623-N1\tablefootmark{c}) & -\\ 
IRAS 16293 A,B & 29 & 19.2 &45 & 0 & 2-4 (A1+A2) & Oph &  52 (A1-A2)  & - \\ 
 && &  &  & 0.2-0.8 (B) &  & 734 (A2-B)  & -\\ 
Oph IRS 63 & 1.2 & 0.7&351 & I & 0.5 & Oph &   $>7000$ \tablefoottext{c}  & \cite{2024FAUSTIRS63DeutDisk}\\ 
L1551 IRS5 N,S & 26.6 & 21.4 & 106 & I & 1-1.7 (N+S)& Tau & 54 (N-S)  & \cite{2020BianchiL1551}\\ 
&&&&&&&& \cite{2025DuranL1551}\\

\hline
\end{tabular}
\tablebib{
   RCrA IRS7B a,b:  \cite{2021SandellFORSCAT,2023OhashieDisk}; RCrA SMM1C, RCrA IRS7B SMM2, RCrA IRS7A: \cite{2021SandellFORSCAT}; RCrA CrA-24: \cite{2015DunhamYoung}; CXO 34: \cite{2011PetersonSpitzer}; Elias 29: \cite{2013GreenEmbedded, 2019OyaSulfur}; VLA 1623 A/B: \cite{2013GreenEmbedded, 2020HsiehVLA1623, 2022OhashiVLA1623}; VLA 1623 W: \cite{2015DunhamYoung,2018MurilloRevised,2023MercimekVLA1623W}, SM1: \cite{2018KawabeDense,2018FriesenALMA} IRAS 16293 A/B: \cite{2015DunhamYoung,2016JorgensenPILS,2020MaureiraOrbits,2018OyaSourceB}; IRS 63: \cite{2008LommenSMA, 2023FloreseDisk}; L1551 IRS5: \cite{2013GreenEmbedded,2024HernandezOrbitsL1551}
   }
\tablefoot{
   Second to last column provides the information about the projected separation to the closest protostar.\\
   \tablefoottext{a}{Observed only at 3mm}
    \tablefoottext{b}{This value is for reference only and comes from a modified blackbody fit to the SED \citep{2018KawabeDense}.  
    \tablefoottext{c}{No other protostar within the ALMA 3 mm field of view.}
    \tablefoottext{d}{\cite{2018ChenOphA}}
    }

   }   
\end{table*}


\section{Sample}
\label{sec:sample}

Protostars in our sample were drawn from the FAUST targets. The only exception is the Class 0 triple system IRAS 16293–2422 (hereafter IRAS 16293) which observations were presented in \cite{2021ZamponiHotdisk} and \cite{2020MaureiraOrbits,2022Maureirahotspots}. We targeted five FAUST fields (RCrA IRS7B, Elias 29, VLA 1623-2417, IRS 63 and L1551 IRS 5) and included in our sample all the Class 0/I sources contained in the respective 3 mm ALMA fields of view. In total, 16 individual disks (7 Class 0, 9 Class I) and 3 circumbinary  disks (2 Class 0, 1 Class I) were observed at 1.3 mm and 3 mm. In the larger 3 mm field view, we detected 4 additional individual disks (1 Class 0, 3 Class I) and 1 circumbinary disk (Class I). Most sources observed at both wavelengths (14/16) have $T_{\rm bol}<200$ K, making the sample more representative of the early protostellar stages. The sampled molecular clouds and adopted distances are Corona Australis ($d=149$ pc, \citealt{2020GalliRCrAdistGaia}), Taurus ($d=146$ pc, \citealt{2019GalliTaurusGaiaVLBI}) and Ophiuchus. For Ophiuchus, we adopted $d=141$ pc \citep{2018DzibRevised} for protostars in the eastern part of the cloud corresponding to IRS 63 and IRAS 16293, and $d=137$ pc \citep{2018OrtizLeonGaia} for the protostellar systems towards Ophiuchus A, corresponding to VLA 1623-2417 system (hereafter VLA 1623), Elias 29 and Oph A SM1. Table~\ref{table:source_prop} summarizes the properties of all the sources presented in this work. All listed properties are presented when available in the literature, with the corresponding references. We include a column with the internal luminosity $L_{\text{int}}$, which is the protostar luminosity without the contribution of UV illumination in the envelope affecting $L_{\text{bol}}$. $L_{\text{int}}$ is obtained from the relation between the flux at 70 $\mu$m and $L_{\text{int}}$, found in \cite{2008DunhamIndentifying} through radiative transfer modelling. The flux at 70 $\mu$m is taken from the eHOPS catalogue (extension of the Herschel Orion protostar survey, or HOPS, Out to 500 ParSecs) updated in \cite{2023PokhreleHOPS}\footnote{ NASA/IPAC Infrared Science Archive, \url{https://irsa.ipac.caltech.edu/data/Herschel/eHOPS/overview.html}}. For the few sources not listed in that catalog (VLA1623 W, OphA SM1, RCrA SMM1C and L1551 IRS 5) the fluxes were obtained from the references provided in Table~\ref{table:source_prop}. We also include a column indicating the projected separation to the nearest protostar within the ALMA 3 mm field of view, providing insight into whether the source is relatively isolated or part of a close multiple system. For reference, the last column lists current FAUST papers showing molecular line emission with a resolution of 50 au towards the individual protostars and fields.



\section{Observations}

\label{sec:data}

The ALMA band 3 and 6 long-baselines observations targeting VLA 1623, RCrA IRS7B, L1551 IRS5, IRS 63 and Elias 29 were taken during 2021 between August and October. The observations were part of the cycle 7 project ID:2019.1.01074.S (PI: Maureira). The most extended configuration C-10 was used for the band 3 observations with baselines ranging from 122 to 16200 m, while C-8 was used for the band 6 observations with baselines ranging from 470 to 11500 m. The configuration and frequency setup was chosen to be combined with the observations from FAUST (PI: S. Yamamoto, ID: 2018.1.01205.L). 
The FAUST observations consisted of two main-array configurations (plus ACA observations for the band 6). Here we only used the most extended configuration for each band, which were taken from October 2018 through March 2020, and corresponded to C-6 for the band 3, while either C-4, C-5 or C-6 for band 6. For the purpose of this project, we refer to the FAUST observations used here as the compact configuration. \\

The spectral setup in band 3 consisted of four spectral windows centered at frequencies of 93.1, 94.9, 107.0, and 105.1 GHz, with a channel width of 488.281 kHz ($\sim$1.5 km s$^{-1}$) and bandwidth of 1.875 GHz. In band 6, the setup consisted of seven spectral windows centered at frequencies of 217.0, 219.9, 219.5, 218.4, 218.2, 231.6 and 233.7 GHz. The narrower spectral windows between 219.9 GHz and 218.2 GHz were set up for lines following the FAUST setup, with a channel width of 122 kHz ($\sim$0.16 km s$^{-1}$) and bandwidth of 58.6 MHz. The rest of the spectral windows had the same channel width ($\sim$0.6 km s$^{-1}$ at 230 GHz) and bandwidth as the band 3 setup. \\

We used CASA \citep{2022CASATeam} to calibrate and image the long baselines observations. Calibration of the raw visibility data was done using the standard pipeline. To create the continuum visibilities we use the continuum channel ranges provided for the pipeline over all the available spectral windows. We checked that the continuum visibilities were not contaminated by line emission. For this, we produced an cleaned image of the continuum visibilities and afterwards appended the model image onto the data visibilities using the ft task in CASA. We then subtracted the model from the data using the CASA task uvsub. Finally, we imaged these residuals checking that only noise was left. For the line rich source L1551 IRS 5 \citep{2020BianchiL1551}, we also created dirty cubes of the visibilities after line flagging and found no obvious remaining lines. Similarly, for the other line rich sources in this study (IRAS 16293-2422 A/B), the line channels were carefully flagged by hand by inspecting dirty cubes of each spectral window \citep{2022Maureirahotspots}. We then averaged in frequency resulting in channel widths of $\sim$10 MHz and $\sim$30 MHz for the band 3 and band 6 data, respectively. These values are conservatively chosen to avoid intensity losses due to bandwidth smearing for sources offset from the phase center. \\


We performed self-calibration separately for the observations in the extended configuration, which were then combined with the self-calibrated compact configuration observations and self-calibrated again together. Details of the calibration and self-calibration procedures for the FAUST observations can be found in \cite{2020BianchiL1551} and \cite{2024MaureiraSM1}. For cleaning and self-calibration we use the CASA tasks tclean and gaincal. The cleaning in the self-calibration iterations was done first with the ``multiscale''  and later with the ``mtmfs'' deconvolver parameter, with the parameter nterms$=$2 for the final phase and amplitude steps. For the parameters ``robust'' and ``uv-taper'' we explored values to match the resolution we obtained from the extended configuration observations imaged with a robust of 0.5. This resulted in typical robust values of 0. A manual mask was set and adjusted during the self-calibration process when necessary. Details of the self-calibration and imaging can be found in Appendix Section~\ref{sec:self-calibration-app}. Table~\ref{table:maps_properties} lists the beam sizes, final rms and imaging parameters of the final continuum images. 

\begin{figure*}[t!]
\centering
   \includegraphics[width=17cm]{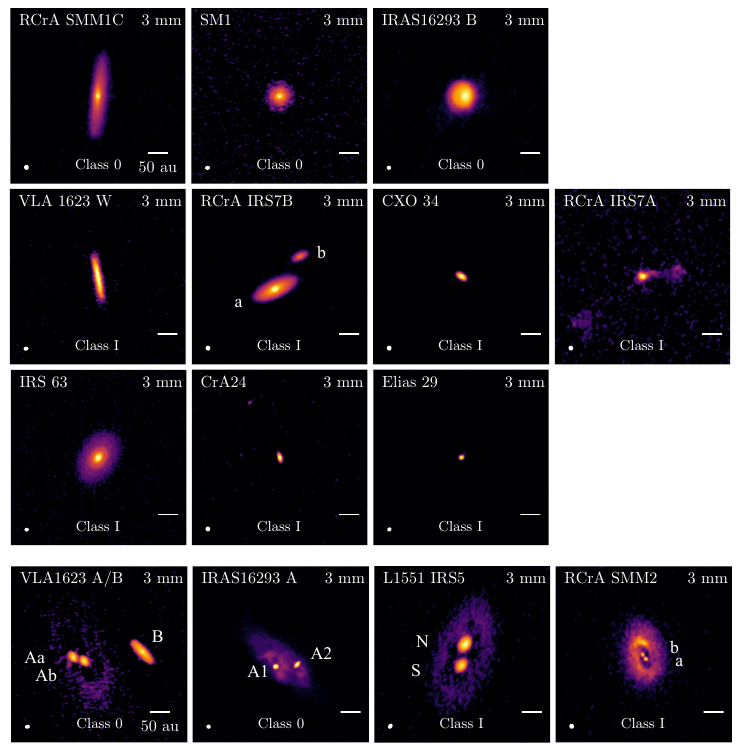}
    \caption{ALMA 3 mm images of all the sources in our sample at the same physical scale. The emission is presented with a log stretch to enhance the weaker extended emission. The images are organized in two groups. The first one (top three rows) correspond to all the systems in which the projected separation to the nearest protostellar neighbor is larger than 100 au, while the second one (bottom row) comprises the systems with a protostellar neighbor below 100 au. Unlike the first group, disk-like circumbinary structures are observed for all sources in the second group. For each group the sources are organized by increasing bolometric temperature. 
       }
         \label{fig:obs_3mm}
\end{figure*}

 \begin{figure*}
\centering
   \includegraphics[width=16.5cm]{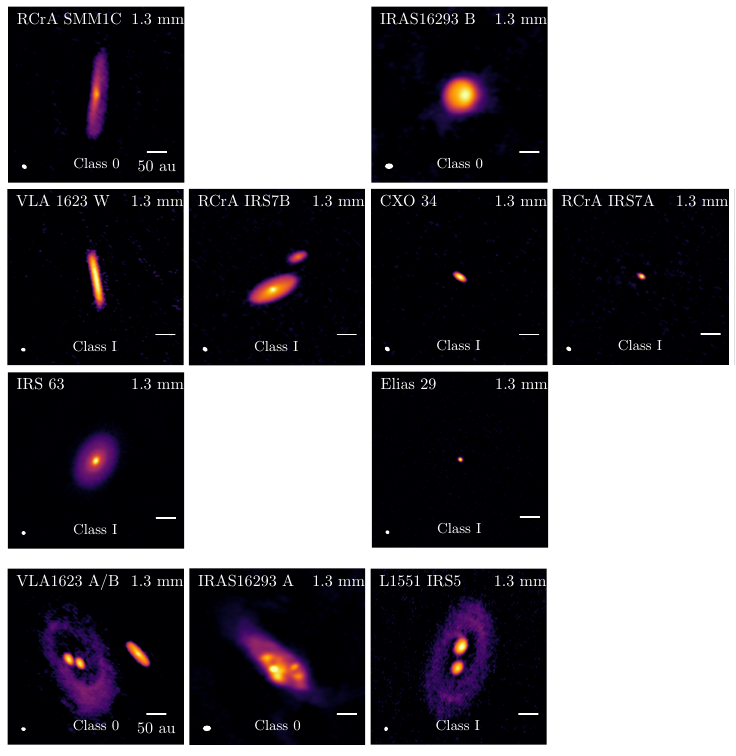}
    \caption{Same as Figure~\ref{fig:obs_3mm}, for the 1.3 mm observations. Some sources detected at 3 mm and not observed at 1.3 mm due to the smaller field of view.}
         \label{fig:obs_1mm}
\end{figure*}

\subsection{Imaging for spectral index}
\label{sec:specindeximaging}

To obtain spectral index maps it is important to create images at the different bands that recover as closely as possible the same spatial scales. Thus we performed additional imaging limiting the baselines so as to cover the same uv-wavelength range in both wavelengths. We use the same tclean setup and scales as during the last self-cal imaging described above, except for the robust parameter. We explored ``uv-taper'' and ``robust'' parameters in tclean to closely match the synthesized beams for the two wavelengths. Finally, we use the restoring beam option in tclean to produce images with the same synthesized beam. The resultant resolution and rms of these images are summarized in Table~\ref{table:maps_properties_foralpha} in the Appendix.

\subsection{Archival data}

We added ALMA band 3/6 observations towards the Class 0 multiple system IRAS 16293-2422 (project ID: 2017.1.01247.S/PI: Dipierro, and  ID: 2016.1.00457.S/PI: Oya) that were observed at similar or slightly lower ($\sim13$ au) resolution. The self-calibration and imaging of the data (including spectral index images) is similar to the procedures described above and can be found in \cite{2021ZamponiHotdisk} and \cite{2022Maureirahotspots}. The resolution and rms of these maps are listed in Tables~\ref{table:maps_properties} and~\ref{table:maps_properties_foralpha}.

\section{Results}
\label{sec:results}

\subsection{Continuum maps}
\label{sec:cont_maps}

 \begin{figure*}
\centering
   \includegraphics[width=15cm]{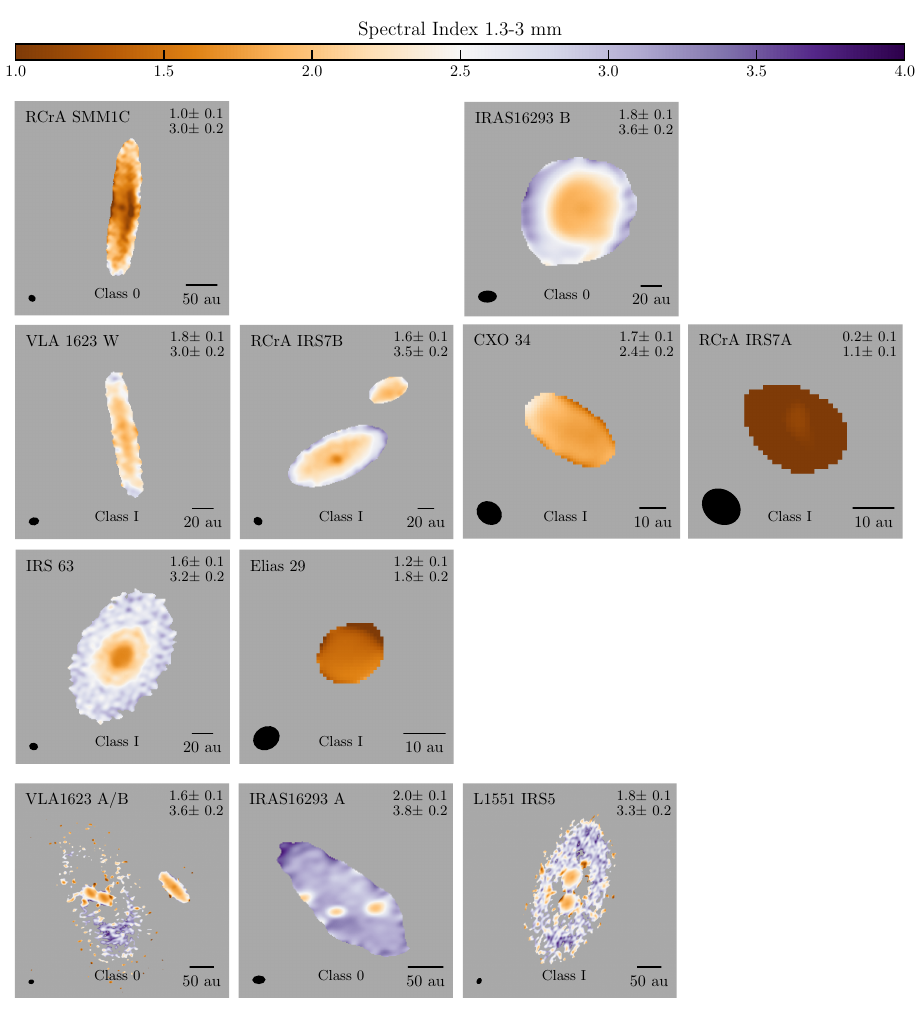}
     \caption{Maps of 1.3-3 mm spectral index for all sources with observations at 1.3 and 3 mm. The ordering is the same as in Figures~\ref{fig:obs_3mm} and~\ref{fig:obs_1mm}. The upper corner of each panel shows the minimum and maximum value within the map, considering only pixels with errors below 0.2 (including both statistical and flux calibration uncertainties).}
         \label{fig:alpha_all}
\end{figure*}

Figures~\ref{fig:obs_3mm} and~\ref{fig:obs_1mm} show a gallery of the continuum images at 3 mm and 1.3 mm for all the sources in our sample, respectively.  The emission is presented with a log stretch to enhance the weaker extended emission. The sources are ordered by increasing bolometric temperature. Multiple protostellar systems with projected separation below 100 au are shown separately in the bottom row. All the sources are presented at the same physical scale. 

Emission associated with disks is observed for all targets. There is a clear diversity of dust disk sizes ranging from a few au (Elias 29) up to $\sim$100 au (SMM1C). Overall, the morphology of the disks appears similar at both wavelengths, except for the Class I protostar RCrA IRS7A. A compact disk structure is detected at 1.3 mm for this source. In contrast, the emission at 3 mm from the central region is more extended and shows a non-axisymmetric shape. The 3 mm observations for this source also show narrow features extending up to $\sim$50-150 au from the disk which are absent at 1.3 mm. The morphology, position angle and 1.3-3 mm spectral index of these elongated features ($\alpha\lesssim-0.2-0.7$) supports a non-dust origin (e.g., free-free), possibly associated with material being ejected from the protostar (see Appendix Section~\ref{sec:irs7a_nondust} for further details). More subtle differences between the 1.3 mm and 3 mm emission, such as asymmetries along the disk major axis, are visually apparent in some sources. These asymmetries are  more pronounced at 1.3 mm than at 3 mm, suggesting they are due to increasing optical depth toward the center or in more inclined disks \citep{2020VillenaveEdgeOnI,2021LinHH212,2021ZamponiHotdisk,2021LiuMagnetically,2023OhashieDisk, 2024Guerra-AlvaradoBand9HLTau, 2024Guerra-AlvaradoIRAS4A1}.

The observations also show emission associated with circumbinary disks (CBDs). These systems were previously resolved as close binaries in ALMA continuum observations by \cite{2018HarrisVLA1623}, \cite{2020MaureiraOrbits} and \cite{2019CruzdeMieraL1551dust} for VLA 1623 A, IRAS 16293 A and L1551 IRS 5, respectively. The Class I RCrA SMM2 is resolved here for the first time into a 14 au separation binary surrounded by a CBD. CBDs are only detected in binaries with projected separations below 100 au. The CBDs show clear inner gaps in all sources, except the Class 0 IRAS 16293 A1/A2, along with azimuthal asymmetries in their brightness distribution more clearly observed at 1.3 mm.

In Table~\ref{table:source_positions}, we summarize the center coordinates, position angles and inclinations for all individual protostellar disks derived from the optically thinner 3 mm emission. The values are obtained from a 2D Gaussian fit to the 3 mm images using the CASA task imfit. Inclinations were derived assuming a circular geometry. The derived inclinations range from 16$^{\circ}$ to 82$^{\circ}$ with a median of 59$^{\circ}$. 

\begin{figure}
  \resizebox{\hsize}{!}{\includegraphics{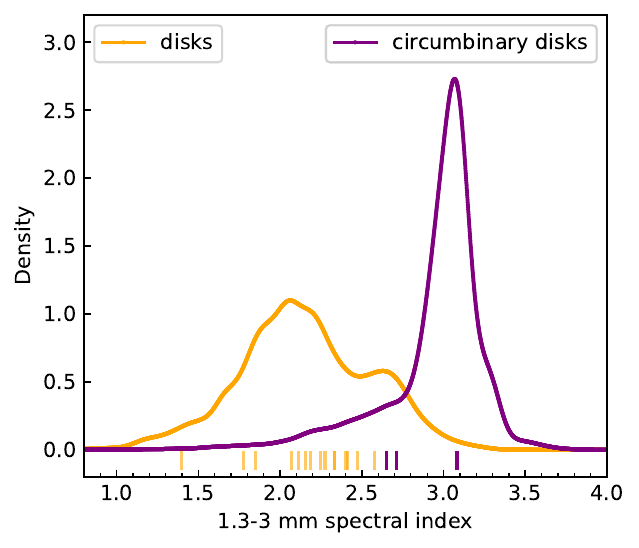}}
   \caption{Spectral index distribution of all disks versus circumbinary material, obtained through a KDE. The vertical lines at the bottom show the mean value of the spectral index for the individual disk sources and circumbinary structures} 
         \label{fig:kde_alpha}
\end{figure}

 \begin{figure*}
\centering
   \includegraphics[width=17cm]{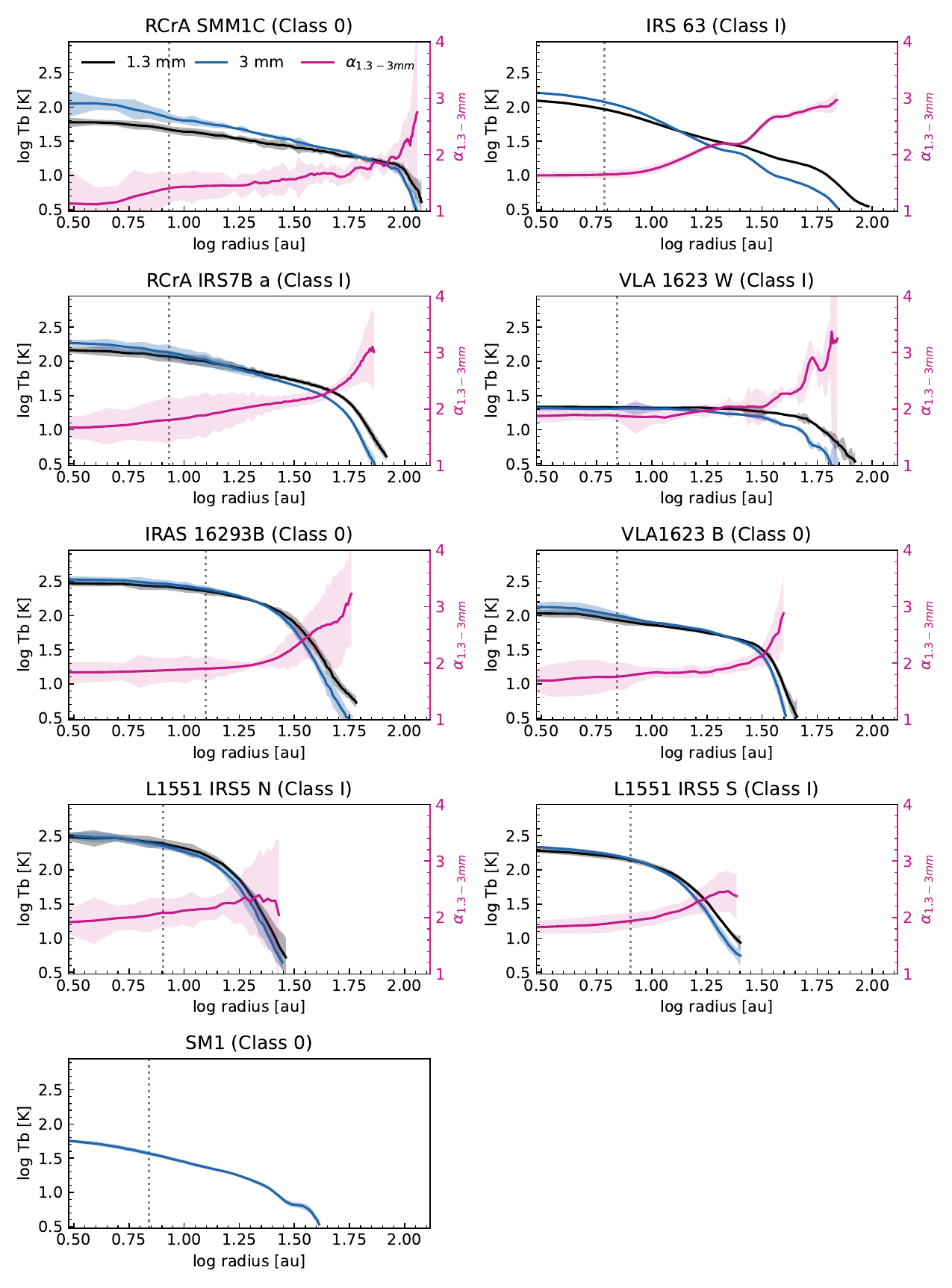}%
    \caption{Intensity profiles on logarithmic scales at 1.3 and 3 mm expressed in brightness temperature, calculated using the full Planck function. The shaded area corresponds to the uncertainty in the mean (Section~\ref{sec:intensity_profiles}). The 1.3-3 mm spectral index $\alpha$ calculated from the profiles is shown in magenta with the corresponding value in the right y-axis. The magenta shaded area shows the uncertainty calculated propagating the errors in the intensity profiles. The vertical dotted line represents the geometric mean of the beam size, indicating that emission beyond this point is spatially well resolved.}
         \label{fig:rad_profiles}
\end{figure*}

\subsection{Spectral Index maps}
\label{sec:spec_index_maps}


We constructed spectral index $\alpha$ maps from the 1.3 mm and 3 mm images built with matching {\it uv}-range and synthesized beam (Section~\ref{sec:specindeximaging}). The spectral index maps were created using
\begin{equation}
\alpha = \frac{\ln(I_{\nu_1}/I_{\nu_2})}{\ln(\nu_1/\nu_2)}, 
\label{eq:spec_index}
\end{equation}
\noindent where $\nu_1$ and $\nu_2$ correspond to 225\footnote{223 GHz in the case of IRAS 16293 A/B}) GHz and 100 GHz, respectively. When examining the maps, some sources showed spectral index maps with systematic gradients along a certain direction, which could be caused by the images at 1.3 mm and 3 mm not being correctly aligned. Small misalignments can arise due to the motion of the source at the different epochs, systematic residual phase errors that are different in different bands or calibrators not being exactly at the same positions at the two frequencies due to opacity effects \citep{2014KutkinCoreShift}. To correct for this, we calculated the shift between the images by looking at their cross-correlation\footnote{This finds the shift that maximizes the similarity between the images}, using the \texttt{scipy} function \texttt{fftconvolve}. The detected shifts range from 1 to 2 pixels which is equivalent to 7.5-15 mas in the case of the Corona Australis field, and 10-20 mas in the case of the Ophiuchus field containing IRAS 16293 A/B. Most of the systematic gradients dissipated after the aligning procedure. See Section~\ref{ap:alignemnt} for further details. The statistical error in $\alpha$ for each pixel is found to be typically below 0.1 for all disks and the CBD IRAS 16293 A, reaching a maximum uncertainty of 0.2 towards the edges. For the CBDs around L1551 IRS5 N-S and VLA 1623 Aa-Ab, statistical uncertainties for each pixel\footnote{ These corresponds to the uncertainties calculated by propagating the rms of the images.} are higher (0.2-0.5). Full maps for the statistical uncertainty of $\alpha$ are in Appendix Figure~\ref{fig:alpha_error_all}. In addition to the statistical uncertainty, all $\alpha$ maps have a systematic 1$\sigma$ error of 0.07 from flux calibration\footnote{The 1$\sigma$ flux calibration error corresponds to 2.5\% and 5\% for band 3 and 6 respectively (see Cycle 9 ALMA Technical handbook, Section 10.2.6).}. \\

Figure~\ref{fig:alpha_all} shows the resultant spectral index maps for the sample. The upper right corner in each panel shows the minimum and maximum value for $\alpha$ in the map. Most disks display widespread low values ($\alpha \approx 2$), with $\alpha$ rising to $\gtrsim3$ towards the edges. While compact disks without clear radial gradients (e.g., Elias 29) may be affected by free-free contamination leading to the observed low $\alpha$ values, sometimes even below 2, similar low $\alpha$ values are also found in larger disks, extending beyond 30 au (e.g., RCrA SMM1C, RCrA IRS7B, VLA1623 W). This implies that free-free contamination, expected only near the central region, cannot fully account for the low observed values. A previously established exception is the compact RCrA IRS7 A disk, where non-dust emission within 10 au yields a mean $\alpha$ of 0.7. \\

Figure~\ref{fig:alpha_all} also shows the 1.3-3mm spectral index for three CBDs. Overall, the $\alpha$ values towards these structures are higher than those found towards the circumstellar disks, either comparable to the values found towards some of the disk edges ($\alpha\approx3$) or larger. In Figure~\ref{fig:kde_alpha}, we compare the distribution of $\alpha$ values in the individual disks versus that towards the CBDs computed using a Gaussian kernel density estimator (KDE) aggregating all pixels in the spectral index maps for all disk and CBD structures. The distributions clearly peak at different values for $\alpha$. For the disks, the median of $\alpha$ is 2.1 with 68\% of the values in the range 1.8-2.6.  For the circumbinary structures, the median value is 3.0, with 68\% of the values in the range 2.7-3.2. The individual disk with the largest average $\alpha$ is the Class I IRS 63. Therein, $\alpha$ values other than those towards the ring structure \citep{2020SeguraCoxFour} and central region, contribute the most to the second peak of the disk density distribution near $\alpha\sim2.6$.\\

\subsection{Radial intensity and spectral index profiles}
\label{sec:intensity_profiles}


We computed azimuthally averaged intensity profiles for the larger disks in the sample. This subsample corresponds to those for which the geometric mean of the deconvolved FWHM axes from the 2D gaussian fit is at least 3 times the size of the synthesized beam in at least one of the wavelengths. In order to compare the intensity at 1.3 mm and 3 mm, the profiles are built from the observations with matching uvrange and synthesized beam in both wavelengths (Section~\ref{sec:specindeximaging}) and after alignment correction (Section~\ref{sec:spec_index_maps}). To build the intensity profile we first deprojected the disks using the center coordinates and inclinations from the 2D Gaussian fit to the 3 mm emission (Table~\ref{table:source_positions}). We then calculated the azimuthally averaged profiles from the deprojected images. For the more inclined disks (RCrA SMM1C, VLA 1623 W, VLA 1623 B) we use only a limited range of azimuthal angles ($<$ 10-20 degrees) around the disk major axis. Similarly, for the circumstellar disk L1551 IRS5 N,  we also excluded some azimuthal angles to the north, to avoid contamination from emission connected to the circumbinary disk (Figure~\ref{fig:obs_1mm}). The uncertainty in each bin is calculated as $\sigma_{std}/\sqrt{N}$, where $\sigma_{std}$ and $N$ are the standard deviation, and number of beams in each bin, respectively. \\

Figure~\ref{fig:rad_profiles} shows the resultant disk profiles in logarithmic scale where the intensity is given as brightness temperature $T_b$ calculated using the full Planck function. All the profiles show a behavior that resembles the classic viscous disk solution \citep{1974LyndenBellEvolution}, with an inner power-law and outer exponential cutoff behaviour. As discussed in \cite{2014BirnstielOuterEdge}, at the outer edges of a disk, where the gas surface density decreases rapidly with radius, the drift speeds are fast for particles of any size. This leads to an early ($\lesssim0.1$ Myr) sharp outer edge in the dust surface density that should be imprinted in the dust continuum emission. Our observations indeed indicate the presence of a sharp outer edge in the dust emission at 1.3 and 3 mm.\\

The profiles at 1.3 mm and 3 mm for each disk resemble each other very closely for the majority of the sources. Another way to look at the similarity of the profiles at 1.3 and 3 mm  is through the spectral index profile, overlaid on the brightness temperature profiles in Figure~\ref{fig:rad_profiles}. The spectral index increases with radius, but the profiles show in more detail that $\alpha$ values in agreement with the optically thick limit of 2 (or below) are observed all the way up to near the location of the disk 'edge' marked by the sudden drop of emission. It is only near this 'edge' where the 3 mm profile shows brightness temperatures below that of 1.3 mm, resulting in $\alpha$ values increasing above 2. Such behavior near and beyond the 'edge' in $\alpha$ is expected when the emission becomes optically thin or at least partially optically thin. The only exception to this behavior is the Class I IRS 63, for which this transition to optically thinner emission occurs at smaller radii than in the rest of the sample. The ring detected in this disk at 1.3 mm  (see also \citealt{2020SeguraCoxFour}) and for the first time here also at 3 mm, falls within this region where $\alpha$ is already above the optically thick limit. Such regions are clearly almost nonexistent for the other disks in the sample. \\

Towards the inner part of the disks, the trend is that $\alpha$ can take values below 2, as seen in the spectral index maps (Figure~\ref{fig:alpha_all}). As expected, this behavior is related to regions for which the 3 mm brightness temperatures is progressively larger than that of 1.3 mm. Such behavior can be expected when the emission is very optically thick at both wavelengths, and can be due to either positive gradients of temperature for increasing depth into the disk (e.g., \citealt{2017LiSpectralIndex,2018GalvanMadridEffects,2021ZamponiHotdisk, 2021LinHH212}) or due to self-scattered emission due to the presence of large grains \citep{2019LiuAnomalously,2019ZhuScattering}.

\section{Analysis}
\label{sec:analysis}

\subsection{Fits to the intensity profiles}
\label{sec:profile_fit}

For more extended disks, for which deprojected intensity profiles were derived (Section~\ref{sec:intensity_profiles}), a 2D Gaussian fit can result in obvious residuals that indicate that this type of intensity profile is not a good model. Thus, in order to better quantify the disk properties and compare them across wavelengths, we fit the deprojected radial intensity profiles for these extended disks with a more general form consisting of two independent power-law indices for constraining the inner and outer cutoff behaviors of the profiles. This form is a variation of the classic solution, and has been used to fit the emission of more evolved Class II disks (e.g., \citealt{2021TazzariMulti}). The intensity profile has the form: 

\begin{equation}
\label{eq:self-similar}
    I(r) \propto  \left(\frac{r}{r_c} \right)^{\gamma_1}\exp\left[-\left(\frac{r}{r_c}\right)^{\gamma_2}\right]
,\end{equation}

\noindent where $r$ is the radius, $r_c$ is the characteristic radius, $\gamma_1$ the inner disk power-law index, and $\gamma_2$ the power-law index describing the slope of the exponential cutoff. Together with $r_c$, $\gamma_1$, and $\gamma_2$, the fourth free parameter is taken as the total flux of the model, $F_{\rm tot}$. Each model was convolved with the synthesized beam, and the values for each of the parameters were obtained using the Python module \texttt{emcee} \citep{2013Foreman-Mackeyemcee}. We refer the reader to \cite{2024MaureiraSM1} for further details of the emcee setup. The resultant fits overlaid in the observations are shown in Appendix Figures~\ref{fig:model_obs_res_plots_b6} and~\ref{fig:model_obs_res_plots_b3}, and the resulting values for the free parameters are listed in Table~\ref{tab:sizes_slopes}.  

For the remaining (more compact disks), we use the results of the 2D Gaussian fit to compute model radial profiles. We use the modified self-similar profile in Equation~\ref{eq:self-similar} with $\gamma_1=0$ and $\gamma_2=2$, resulting in a Gaussian profile. We compute $r_c$ using the value for the deconvolved FWHM (major axis) as FWHM/$(2\sqrt{\ln2})$. The final parameter $F_{\rm tot}$, is computed as the total flux from the 2D Gaussian fit divided by $\cos(i)$, which corresponds to the face-on flux, i.e., the same quantity computed from the modified self-similar profile fit to the extended disks. Although for an optically thin disk the total flux does not change with inclination, the $\cos(i)$ correction applies for optically thick emission \citep{2021TazzariMulti} which is already suggested by the observed low $\alpha$ values (Figure~\ref{fig:alpha_all}). We will be further discussing the optically thick nature of the emission throughout the article.

\subsection{Disk sizes at 1.3 and 3 mm}
\label{sec:disk_sizes}

\begin{figure}
  \resizebox{0.9\hsize}{!}{\includegraphics{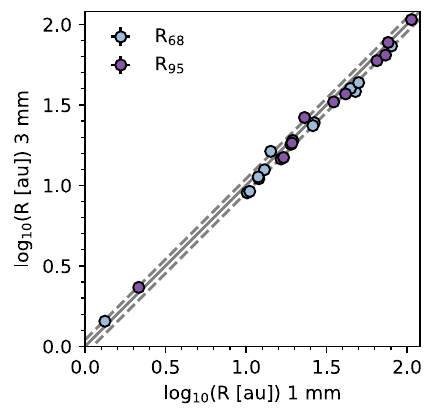}}
   \caption{Comparison of disk radii, measured as R$_{68}$ or R$_{95}$, between 1.3 and 3 mm. The solid line shows the 1:1 ratio, while the dashed lines show a 10\% difference.}
         \label{fig:size_comp_1-3mm}
\end{figure}

\begin{figure}
  \resizebox{\hsize}{!}{\includegraphics{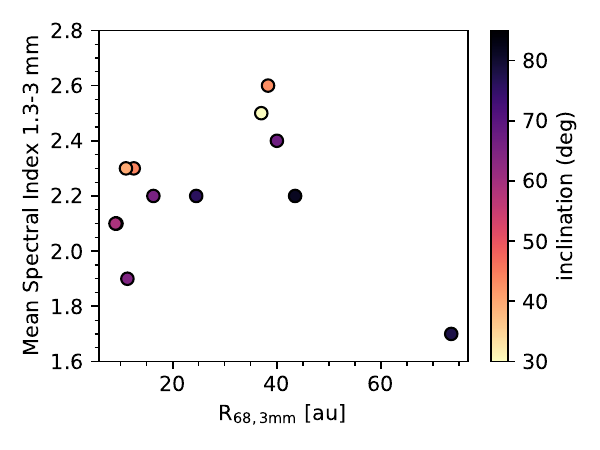}}
  \caption{Mean spectral index for the Class 0/I disks in the sample as a function of the radius enclosing 68\% of the 3 mm flux. The color of the symbols indicate the disk inclination. }
         \label{fig:alpha_vs_size}
\end{figure}

\begin{figure}
  \resizebox{0.88\hsize}{!}{\includegraphics{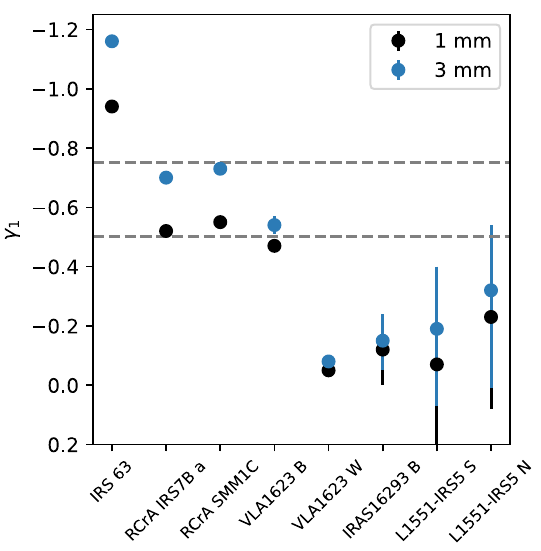}}
  \caption{Index of the inner disk power-law $\gamma_1$ obtained from the fit to the radial intensity profiles (see Section~\ref{sec:profile_fit} for details). The dashed lines mark the index values of -0.5 and -0.75, indicative of passive and viscous heating, respectively. }
         \label{fig:gamma1_sources}
\end{figure}

\begin{figure*}
\centering
   \includegraphics[width=17cm]{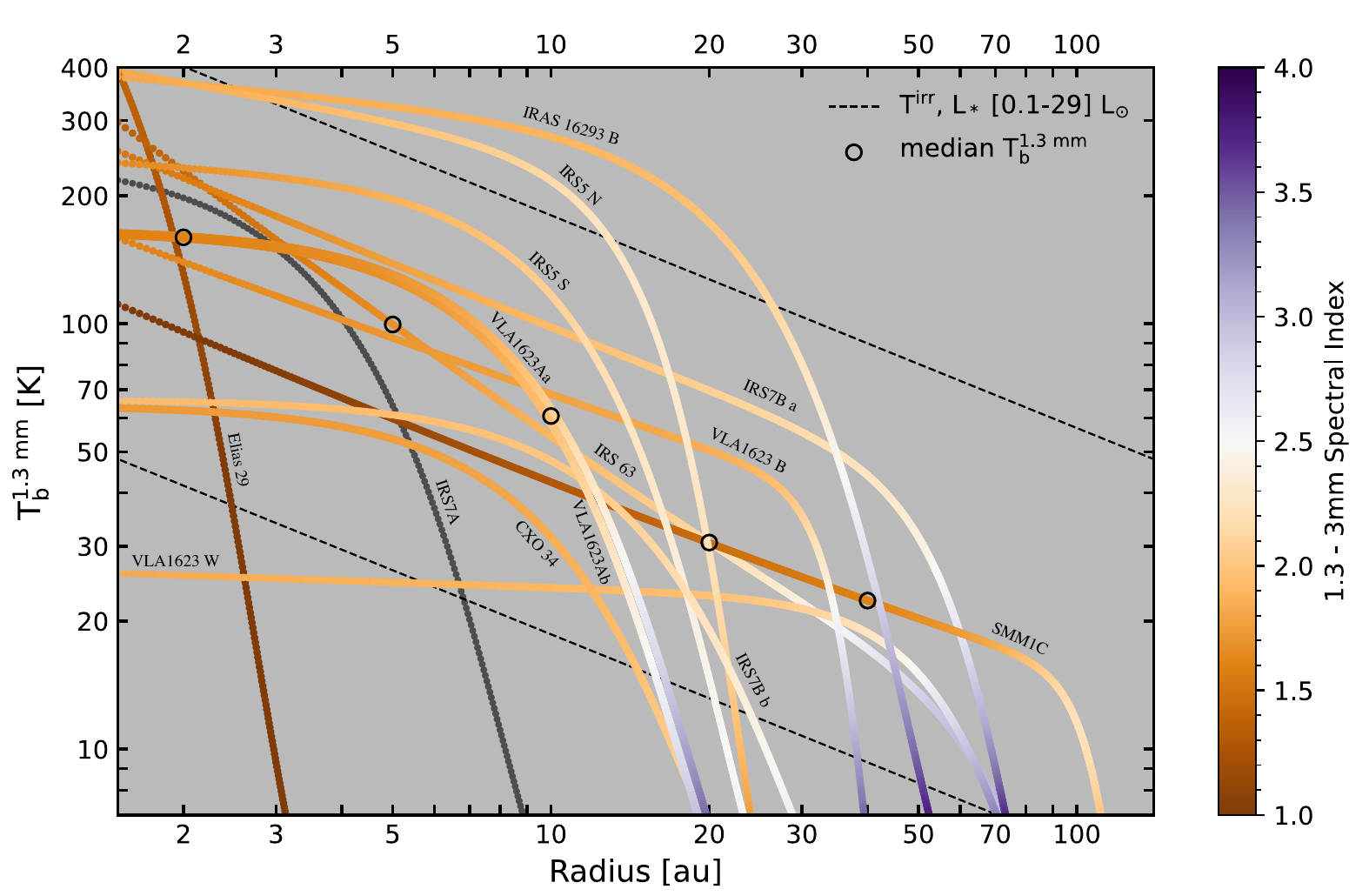}
    \caption{Intensity profiles at 1.3 mm for all disks resulting from either a Gaussian fit (for the most compact disks) or a modified self-similar profile fit (Section~\ref{sec:profile_fit}). The colors of the curves correspond to the resultant 1.3-3 mm spectral index at each radii, except RCrA IRS7A due to the non-dust emission at 3 mm (Section~\ref{sec:irs7a_nondust}). The circles show the median value for the brightness temperature at certain radii. The dashed lines correspond to the theoretical approximation for the midplane temperature due to irradiation ($T\propto r^{-0.5}$) considering the range of bolometric luminosities of the sample (see Section~\ref{sec:tb_vs_td_alpha} for details).}
         \label{fig:Tb_alpha_1.3mm}
\end{figure*}

In the literature, disk sizes are often defined as the radius containing a certain fraction of the total flux. This allows better comparison across studies using different prescriptions to fit the disk emission \citep{2023MiotelloPPVII}. Typically, the reported sizes correspond to R$_{68}$ or R$_{95}$ corresponding to the radius containing 68\% and 95\% of the total flux, respectively. In principle, R$_{95}$ is a more robust measurement of the disk, but given its dependence on the fainter and rapidly decaying outer emission, its use is more limited in the literature. Tables~\ref{tab:sizes_slopes} and~\ref{tab:sizes_slopes_gaussian}  summarize the resultant R$_{68}$ and R$_{95}$ values derived from the modified self-similar profile and the Gaussian fit, respectively. \\

Figure~\ref{fig:size_comp_1-3mm} shows the comparison of the disk sizes at 1.3 and 3 mm using R$_{68}$, as well as R$_{95}$. In both cases, the sizes are almost the same for all disks in the sample. On average, the R$_{68}$ radii are only 7\% smaller at 3 mm compared to 1.3 mm. The agreement between R$_{95}$ at 1.3 and 3 mm is even better, with an average difference of only 3\%. We note that similarly small differences (below 9\%) were measured for R$_{68}$ at 0.9, 1.3 and 3 mm, for a sample of Class II disks in Lupus \cite{2021TazzariMulti}.

\subsection{Spectral index versus disk size}
\label{sec:alpha_vs_size}


Figure~\ref{fig:alpha_vs_size} shows the mean spectral index value for the individual disks in our sample as a function of R$_{68}$ at 3 mm. In order to also investigate possible trends with inclination, we exclude disks for which the spectral index map for the disk is not well resolved\footnote{In these cases, $R_{95}$ values are smaller than the synthesized beam at one or both wavelengths. IRS7A also has strong evidence of non-dust emission at 3 mm.} (RCrA IRS7A, Elias 29 and IRAS 16293 A1/A2). There is a trend of increasing spectral index with radius, in agreement with what is observed for Class II disks  \citep{2021TazzariMulti,2024ChungClassII}. Likewise, there is a trend with inclination such that for a given size, the mean spectral index is lower for more inclined disks. Such a trend is expected when significant optical depth is playing a role in the values for $\alpha$. This adds scatter to the relation and can help explaining why the largest disk in our sample ($i\sim78^{\circ}$) drops out of the trend of increasing $\alpha$ with size. Similar scatter and drops from the correlation are observed for Class II disks \citep{2021TazzariMulti,2024ChungClassII}.

\begin{figure*}
\centering
   \includegraphics[width=17cm]{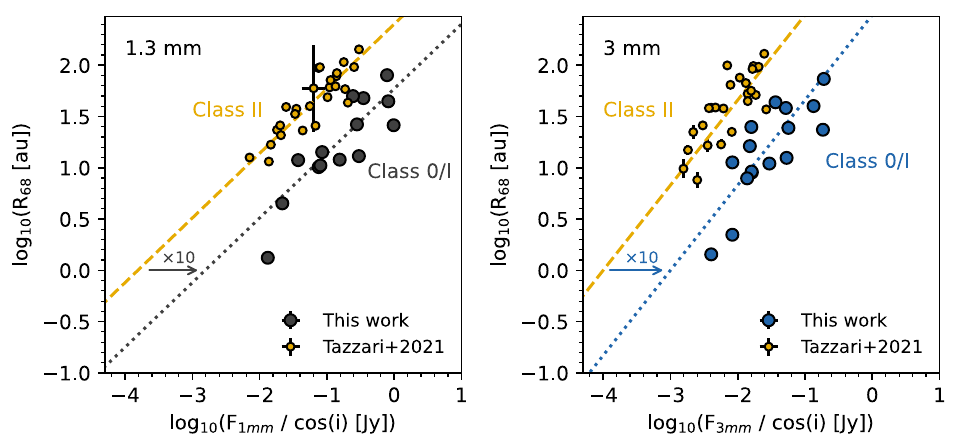}
     \caption{Size and luminosity relations at 1.3 and 3 mm (circles). Left: R$_{68}$ vs flux at 1.3 mm. Right: R$_{68}$ vs flux at 3 mm. Fluxes in both panels have a correction for inclination which is valid for optically thick emission. The dashed lines show the linear regression results for Class II sources at the two wavelengths, derived in \cite{2021TazzariMulti}, which also include the correction for optically thick emission. The dotted lines corresponds to the relation for Class II considering a shift in the fluxes of a factor of 10.}
         \label{fig:size_lum}
\end{figure*}

\subsection{Inner slope at 1.3 and 3 mm}
\label{sec:inner_slope}

Figure~\ref{fig:gamma1_sources} shows the resultant slopes for the inner part of the disks from the modified self-similar fit measured at 1.3 mm and 3 mm. The results show a range of power-law index values ranging from $\sim-1.2$ to $\sim0$. The values are similar at both wavelengths, in agreement with the similarity between the profiles. There is a trend for the values measured at 3 mm to be higher than the one measured at 1.3 mm for each source such that on average the index at 1.3 mm is 24\% shallower than at 3 mm. 



\subsection{Comparison of brightness and dust temperature}
\label{sec:tb_vs_td_alpha}

Figure~\ref{fig:Tb_alpha_1.3mm} shows all the disk intensity profiles at 1.3 mm expressed in brightness temperature $T_b$, using the results from the model radial profile fits (Section~\ref{sec:profile_fit}). At each radius, the color of the curves tracks the resulting model spectral index between 1.3 and 3 mm. The median $T_b$ of the fit profiles at 1 au is $\sim$ 170 K. At 5 au, where the profiles still show $\alpha\approx2$ values, the median is $\sim$ 100 K. At a distance of 40 au, where most of the disks still show $\alpha\approx2$ values, the median brightness temperature is $\sim23$ K. For comparison, the dashed lines in Figure~\ref{fig:Tb_alpha_1.3mm} correspond to minimum and maximum midplane temperature profiles based on the lowest and highest $L_{\text{bol}}$ in the sample. The midplane temperature as a function of radius is calculated as 
\begin{equation}
T^{\mathrm{irr}}_{mid}=\left(\frac{\varphi L_{*}}{8\pi r^2\sigma_{SB}}\right)^{0.25}
\label{eq:Tmid_standard}
\end{equation}

\noindent where $\sigma_{SB}$ is the Stefan-Boltzmann constant. The above expression is an approximation valid for passively heated disks \citep{1997ChiangSpectral}.
To provide a lower and upper limit to the sample midplane temperature, we consider the limiting values 0.1 and 29 L$_{\odot}$ for the protostar luminosity $L_{*}$, and 0.01 and 0.3 for the flaring angle\footnote{We note that considering a radial dependence on the flaring angle according to equation 5 in \citep{1997ChiangSpectral} changes the temperature only by about 10\%.} $\varphi$. The values for $L_{*}$ are taken from the range of $L_{\text{bol}}$ of the sample (Table~\ref{table:source_prop}). These lower and upper limits to $L_{*}$ and $\varphi$ correspond to H/R between $\sim$0.05 and $\sim$0.15 for a 1 M$_{\odot}$ protostar. We observe that the disk $T_b$ profiles as well as median $T_b$ values for the sample cover the same order of magnitude as the predicted range of dust midplane temperatures for irradiated disks, which does not change if we consider instead $L_{*}=L_{int}$. This supports the interpretation of $\alpha\approx2$ as due to optically thick emission. We also note that the observed $T_b$ values for at least two disks (IRAS 16293 B and L1551 IRS 5N) are above the derived upper limit for the temperature due to irradiation. This is in agreement with a previous study for IRAS 16293 B in which mechanical heating (dissipation due to accretion/viscosity) was needed to explain the high fluxes \citep{2021ZamponiHotdisk}. Similarly, L1551 IRS 5 is a FU Ori like source \citep{1985MundtOpticalL1551}, for which viscous heating can also be expected. The comparison between the observed $T_b$ and the predicted values due to irradiation for each individual disk will be discussed in detail in Section~\ref{sec:temp_slope}.\\



\subsection{Disk size-luminosity relations}

Figure~\ref{fig:size_lum} shows R$_{68}$ versus the disk flux at 1.3 and 3 mm, and compares the observed Class 0/I disks with the relations derived for Class II disks in \cite{2021TazzariMulti}. The relation is shown as a function of the flux of the disk as seen face-on, and all fluxes are rescaled at a common distance of 150 pc. The inclination correction for the face-on flux for optically thick emission (Section~\ref{sec:profile_fit}) was also applied by \cite{2021TazzariMulti} for the sample of Class II disks, motivated by the non-negligible optically thick fractions found throughout the sample, and an improvement in the tightness of the relation after applying the correction \citep{2021TazzariMulti}, which also occurs in our sample. The comparison reveals that for the observed Class 0/I disks there is also a correlation between R$_{68}$ and the disk flux at 1.3 and 3 mm. Although the slopes appear to be the same, it is clear that the linear relation for the Class 0, and I disks in our sample is shifted from the one derived for the Class II sources, even after considering the observed scatter in the Class II relation \cite{2021TazzariMulti}. For a given size, the fluxes for the Class 0/I disks in our sample are $\sim10\times$ higher.

\section{Discussion}
\label{sec:discussion}

\subsection{Optically thick emission and comparison with Class II disks}
\label{sec:disc_optdepth_size_luminosity}

\begin{figure*}
\centering
   \includegraphics[width=15cm]{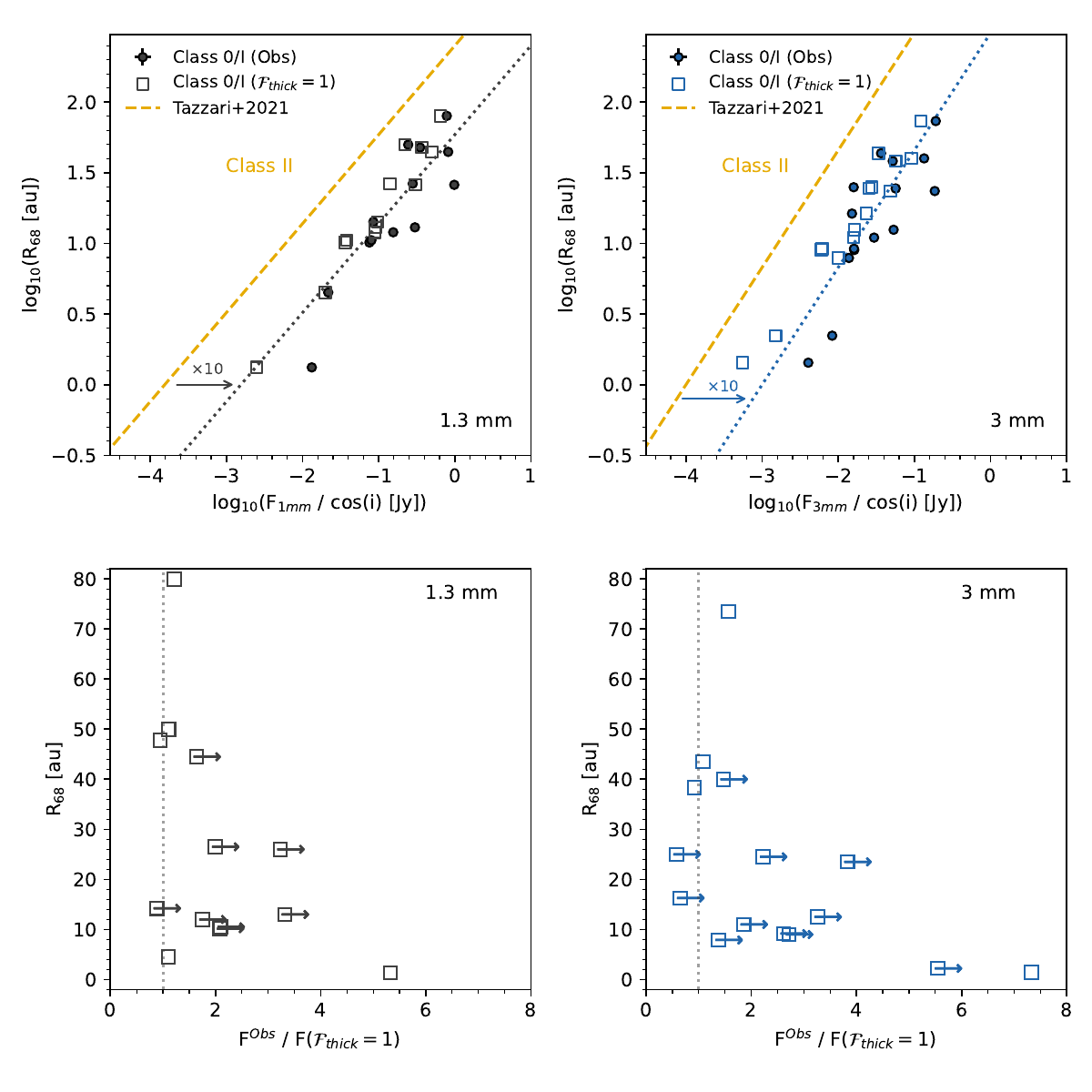}
     \caption{Comparison between the observed disk luminosity and the predicted values assuming the disks are fully optically thick ($\mathcal{F}_{thick}=1$). Top: Same as Figure~\ref{fig:size_lum}. Filled circles are the observations and squares are the predicted luminosity for each disk if the emission is fully optically thick. Bottom: Size versus the ratio of the observed disk luminosity to the predicted one assuming the entire disk is optically thick. Lower limits for the ratio are due to the assumed internal luminosity being an upper limit.}
         \label{fig:size_lum_constalpha}
\end{figure*}

The observations reveal that the observed sample of Class 0/I disks, encompassing different molecular clouds, multiplicity, and even dust disk sizes, have significant optically thick emission at 1.3 and 3 mm. This is supported by (a) the low spectral index values for the disks with a median of $\alpha=2.1^{+0.6}_{-0.2}$, along with per disk median values all below $\alpha\lesssim$2.6, (b) that the median $\alpha$ value for a given size tends to be lower for higher inclinations, and (c) the elevated brightness temperatures with values in agreement (or even above) the predicted temperatures due to irradiation. Results (a) and (b) are very similar to those found towards Class II disks. Low values for the spectral index ($\sim$2) have also been measured at different submm/mm wavelengths for large samples of Class II disks across different regions (\citealt{2021Tazzari3mmLupus, 2024ChungClassII,2025GarufiCentSED,2025PainterDenseSED}). In addition, \cite{2021TazzariMulti}, \cite{2024ChungClassII} and \cite{2025GarufiCentSED} also find that the average per disk spectral index increases with disk size in a similar fashion as our observations of Class 0 and I disks. 

However, result (c) shows differences between the Class 0/I sample and observations of Class II disks. The brightness temperatures of Class II disks appear to be lower than those of the Class 0 and I disks in this work. For instance, \cite{2018AndrewsScalingRelations} measures a median of $\sim$5 K at 40 au at 0.98 mm, while the median of our sample at that distance is about 5$\times$ higher ($\sim$23 K) at 1.3 mm (Figure~\ref{fig:Tb_alpha_1.3mm}). Likewise, similarly low values (T$^{1.3mm}_b\sim$1-10 K) are observed at 40 au in the DSHARP sample \citep{2018HuangDSHARPII}. These low observed brightness temperatures in the Class II disks have led some studies to conclude that most of the disk emission is optically thin at these wavelengths, even for disks observed at high-resolution (e.g., \citealt{2018HuangDSHARPII}). However, low brightness temperatures can also be explained if the emission is actually optically thick and dust scattering opacities are significant \citep{2019LiuAnomalously,2019ZhuScattering}, or if there are unresolved optically thick structures with a filling factor below 1, for instance optically thick rings with optically thin gaps \citep{2012RicciEffect,2018AndrewsScalingRelations}. The latter scenarios are indeed in agreement with multi-wavelength observations of Class II disks mapped with a resolution comparable to the one presented here \citep{2019ZhuScattering,2019CarrascoRadial,2022GuidiDistribution}. 

In the case of larger samples of Class II disks, for which the resolution of the observations do not allow to probe the presence of substructures, it has been shown that the size-luminosity relation for Class II disks can be explained if less than half of the flux was optically thick \citep{2018AndrewsScalingRelations}. They discuss that  higher fractions made the disks too luminous for a given size. Since we showed that the Class 0 and I disks in our sample are brighter for a given size that Class II disks, here we discuss whether this difference can be explained with a higher fraction of the Class 0/I disk's area being optically thick, in line with the resolved dust emission and spectral index maps presented in this work. To test this idea, we followed \cite{2018AndrewsScalingRelations} and compare the observed location of the Class 0/I disks in the size luminosity relation with the predicted location assuming the disks are fully optically thick. To do a simple first comparison, we followed \cite{2018AndrewsScalingRelations} and assume a temperature radial profile given by irradiation with a power law in radius $r$ and protostar luminosity $L_*=L_{int}$. The results in \cite{2018AndrewsScalingRelations} can be explained using $T=T_0(r/r_0)^{-0.5}(L_*/L_{\odot})^{0.25}$ with $T_0=30$ K at $r_0=10$ au for the temperature parametrization. This comes from the approximation of midplane temperature in passive disks where $T\propto(\varphi L_*/r^2)^{
0.25}$, assuming a constant flaring angle $\varphi$ \citep{1997ChiangSpectral}. We then assume a radial intensity profile given by:

\begin{equation}
I_{\nu} = \left\{
\begin{array}{ll}
      \mathcal{F}_{thick}\cdot B_{\nu}(T) & r\leq R_0 \\
      0 & r> R_0 \\
   
\end{array} 
\right.
\end{equation}

\noindent where $\mathcal{F}_{thick}$ is a constant that takes values between 0 and 1, and thus reflects the fraction of the disk flux that is optically thick. Following \cite{2018AndrewsScalingRelations}, $R_0$ is taken as the characteristic radius $r_c$, as beyond this radius typically less than 10-20\% of the total flux is emitted. In cases where a larger percentage of the flux is outside $r_c$, we set $R_0=R_{90}$. For single protostars we use directly  $L_*=L_{int}$ with the values in Table~\ref{table:source_prop}. For close multiples for which we only have $L_{\text{int}}$ values for the entire system, we assume that each member has the same $L_*$, and equal to the one measured for the entire system. Thus, for those disks, the predicted optically thick disk luminosities are a very conservative upper limit. As a check, we showed that if the sample of Class 0/I disks were to have $\mathcal{F}_{thick}=0.3$ the predicted location in the size luminosity relation would match that of the Class II disks (see Figure~\ref{fig:size_lum_f0.3} in the Appendix), thus supporting the methodology and the need for a higher optically thick fraction to be able to reach the higher luminosities observed for the younger disks. 

Figure~\ref{fig:size_lum_constalpha} top panels show the predicted location of the Class 0/I disks if we set $\mathcal{F}_{thick}=1$, i.e. fully optically thick disks. The predicted location is marked with empty squares, while the observed location is marked with filled circles. The predicted location more or less matches the observed location for some of the disks, but there are several disks that remain too luminous. To zoom into the difference between predicted and observed luminosity, the bottom panels in Figure~\ref{fig:size_lum_constalpha} show the ratio between the observed disk flux and the one predicted using $\mathcal{F}_{thick}=1$ for each disk as a function of size. Observed fluxes that are at least 2$\times$ higher than what is predicted by fully optically thick emission are observed in 6 out of 13 disks at 1.3 mm, and 7 out of 15 disks at 3 mm, thus $\sim$46\% at both wavelengths. This percentage is really a lower limit considering that for several disks, those in closer multiple systems, the ratio is only a lower limit given that we have assigned the luminosity of the system to each member. The agreement between the predicted and observed luminosity seems to be size-dependent such that the luminosities are matched for the larger disks ($R_{68}\gtrsim$40-50 au), while discrepancies are more apparent on the more compact disks. We checked if a flaring might be the reason for the discrepancies and find that while a non-constant flaring does help, it cannot not remove all the discrepancies (see Appendix~\ref{sec:alternative_temp_profile} for details). Assuming $L_*=L_{bol}$ along with a radius-dependent flaring still does not completely resolve the discrepancies for IRAS 16293B, Elias 29, and L1551 IRS5N disks (see Figure~\ref{fig:size_lum_varalpha_lbol}), which is a conservative number of sources given the lower limits due to the assumed luminosities in multiple systems. 

A factor of 2 or more is difficult to explain by possible errors in $L_*$ or flaring as it would require at least a factor of $2^{1/0.25}=16$ higher than $L_*$ or flaring, assuming Rayleigh Jeans and the temperature prescription in Equation~\ref{eq:Tmid_standard}. An extreme case is the single Class I disk Elias 29, the smallest disk in our sample. At 1.3 mm, this disks is from 2 to 5$\times$ brighter than its fully optically thick prediction, considering all above temperature parametrizations. The results suggest that the temperatures in these disks are higher than what is expected from irradiation alone, and additional heating such as viscous heating might need to be considered. As this type of heating is expected to dominate at smaller radii, it would also naturally explain why smaller disks show the higher discrepancies between the predicted and observed fluxes.

\begin{figure*}[t!]
\centering
   \includegraphics[width=13cm]{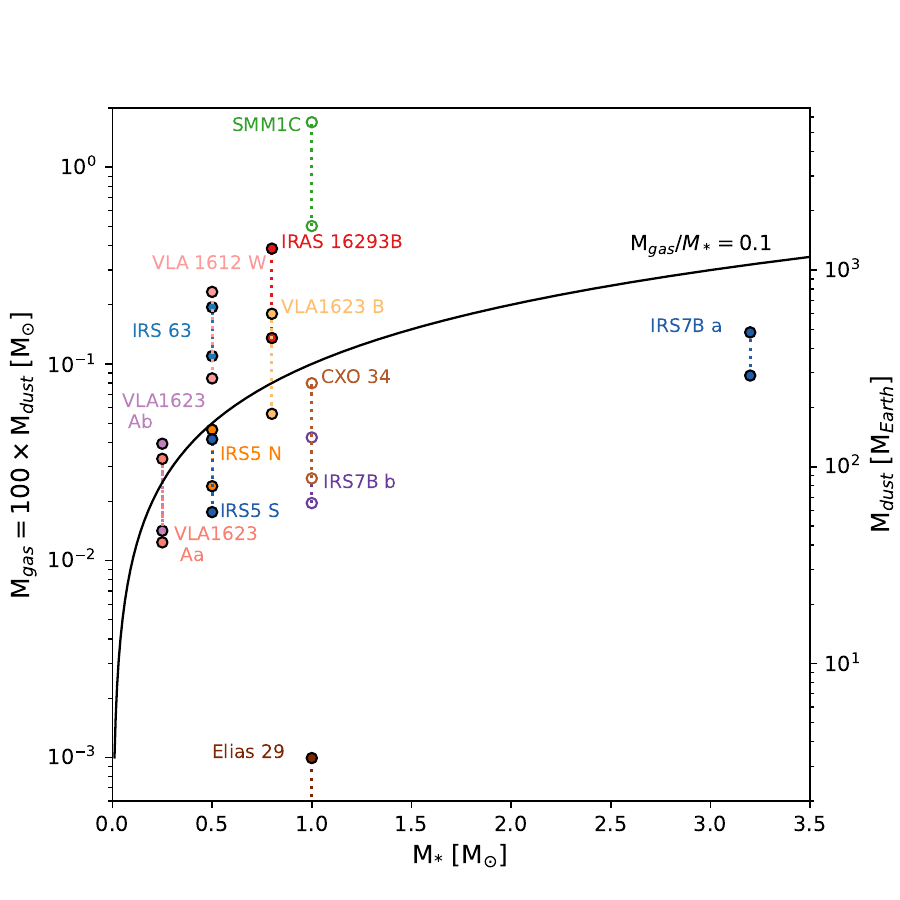}
      \caption{Estimated disk masses versus the protostar mass. The right y-axis corresponds to the dust mass while the left y-axis corresponds to gas mass calculated assuming a gas-to-dust ratio of 100. The solid line represents a criterion for gravitational instability such that disks above the line are unstable. Empty symbols mark sources with no estimates in the literature of the stellar mass for which we assume M$_*=1$ M$_{\odot}$. For close binaries for which only the combined protostar mass is reported in the literature we assigned half that mass to each component. }
         \label{fig:Mdisk_Mstar}
\end{figure*}

 \begin{figure*}[t!]
\centering
   \includegraphics[width=17cm]{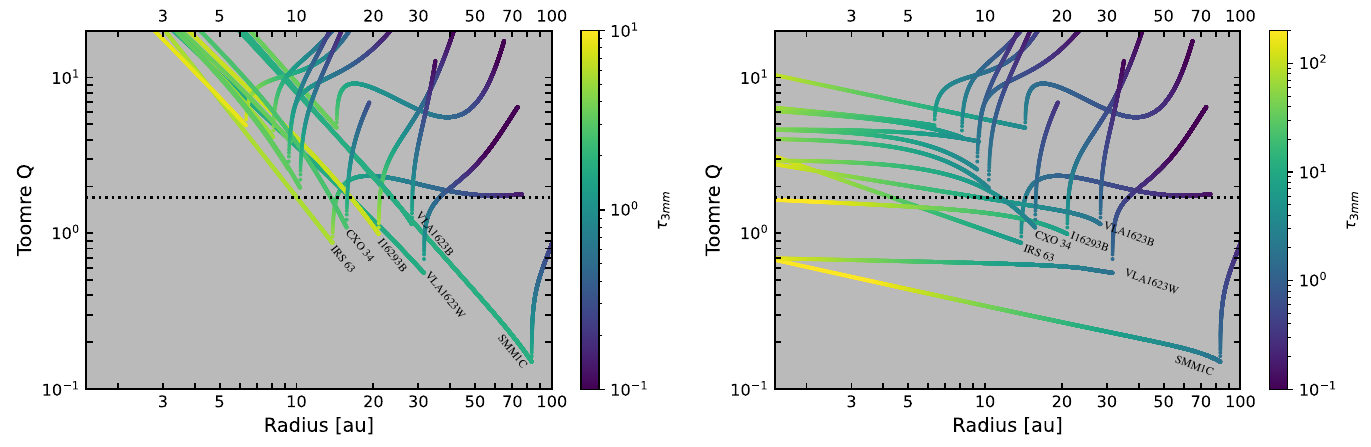}
      \caption{Profiles of the Toomre Q parameter as a function of radius. The left panel shows the case where we extrapolate the optical depth to remain constant inwards of the radius for which $\alpha\approx2$, while the right panel assumes the optical depth increases as $r^{-1.5}$ therein, thus providing an upper and lower limit for Q, respectively. The curves are colored by the resultant dust optical depth at 3 mm. The horizontal dotted line corresponds to $Q=1.7$ as a reference value below which the disk can develop gravitational instabilities.
      See Section~\ref{sec:disk_masses} for more details. }
         \label{fig:toomre_q}
\end{figure*}


\subsection{Optical depths at 1.3 and 3 mm}
\label{sec:opth_depths}

Here we attempt to quantify and discuss the optical depth without the need for $L_*$. Using the modeled disks intensity profiles\footnote{For the fit obtained from emcee, we use the median of the posterior solutions as the model.} and resultant spectral index (Figure~\ref{fig:Tb_alpha_1.3mm}), we can compute a profile for the optical depth at 3 mm $\tau_{3mm}$. For this, we use a modified black-body for the intensity profile $I_{\nu}\sim B_{\nu}(T)(1-e^{-\tau_{\nu}})$. For the temperature, we assume it to be equal to the observed brightness temperature until $r(\alpha=2)$. For larger radii we extrapolated this value assuming a $r^{-0.5}$ profile. We find that all disks, except IRS 63, have at least 70\% (and up to near 100\%) of their 3 mm flux emitted from regions with $\tau_{3mm}\gtrsim1$. The median extent of the $\tau_{3mm}\gtrsim1$ region is the disk characteristic radius $r_c$. These values are just a lower limit at 1.3 mm. Assuming $\beta\sim1$ (dust emissivity index) the sample has a median of $\sim87\%$ of the disk flux with $\tau_{1.3mm}$ above 1, in line with the previous discussion for which we assumed a temperature profile based on $L_{\text{int}}$ (Section~\ref{sec:disc_optdepth_size_luminosity}). These results are summarized in Appendix Figure~\ref{fig:opt_thick_frac}. 

\subsubsection{Disk masses and dynamical state}
\label{sec:disk_masses}

To provide estimates of the disk masses, we use the above derived optical depth profiles at 3 mm, extrapolating inwards from the maximum derived near $r(\alpha=2)$ considering a $r^p$ profile with $p=0$ and $p=-1.5$, as a lower and higher mass limiting cases, respectively. Then, the mass of the disk is calculated as $M_{disk} = 2\pi \int_0^{R} \Sigma(r) r dr,$
where $\Sigma(r) =\tau_{3mm}(r)/\kappa_{3mm}$ and $\kappa_{3mm}$ is the dust opacity, and $R=r_c$. We used an opacity at 3 mm of $\sim$ 1 cm$^2$ g$^{-1}$ \citep{1990beckwithSurvey, 2018BirnstielDSHARP}. Figure~\ref{fig:Mdisk_Mstar} shows the estimated range of disk masses in solids and for the gas assuming a gas-to-dust ratio of 100. The mass in solids spans four orders of magnitude, from $\sim2$ M$_{\oplus}$ (Elias 29, $R_{95}\approx2$ au) up to a few times $10^3$ M$_{\oplus}$ (RCrA SMM1C, $R_{95}\approx100$ au). Most of the disk dust masses are in the range 30-900  M$_{\oplus}$, corresponding to gas masses in the range 0.01-0.3 M$_{\odot}$. The values are in agreement with the estimates in \cite{2020TychoniecDust} for Class 0/I disks in Perseus using observations at 9 mm assuming optically thin conditions, with an opacity value consistent with the one we used at 3 mm if extrapolated using $\beta\sim1$. The derived range of Class 0/I disk masses in Perseus was found to be comparable or above the masses in observed exoplanets \citep{2020TychoniecDust}, thus a similar conclusion can be applied to the masses derived here. The results show that these relatively high masses, required to explain exoplanetary systems, are not unique to a particular cloud, and that it is important to consider a high fraction of optically thick emission if masses are estimated for young disks from continuum ALMA observations, even at 3 mm. Compared with numerical simulations, the estimated masses also agree well with the median disk gas mass of $\sim$0.04-0.06 M$_{\odot}$ found in the synthetic populations in \cite{2018BateDiversity} and \cite{2021LebruillyProto}.

We caution that an important source of uncertainty in this mass estimation, as well as others in the literature, is the value for the dust opacity at a particular wavelength, which depends on the dust composition, morphology and distribution of grain sizes \citep{2014TestiDustEvolution,2023MiotelloPPVII,2024BirnstielReview,2024LiuPorousDust}. As a reference for some of the most significant variations, the estimated masses would increase by a factor of 2–3 if we adopted DSHARP opacities with a maximum grain size of $a_{max}\sim1$ mm \cite{2018BirnstielDSHARP}, \cite{2010RicciDust} opacities with $a_{max}\lesssim100$ $\mu$m, or DIANA opacities with $a_{max}\lesssim300$ $\mu$m \citep{2016WoitkeDIANA}. Conversely, the masses would decrease by approximately a factor of 2 if we used \cite{2010RicciDust} opacities with $a_{max}\sim1$ mm. Another poorly constrained source of uncertainty in observations of young Class 0/I disks are possible changes of the gas-to-dust ratio \citep{2020LebreuillyDustRichDisks,2023MiotelloPPVII,2023OhashiDustEnrichment}. \\

The derived mass range in Figure~\ref{fig:Mdisk_Mstar}, considering the above uncertainties, also agrees better with Class 0/I disk mass estimates in Orion from single-wavelength ($\lambda\lesssim1.3$ mm) observations that are based on marginally gravitationally unstable disk models \citep{2022XuTesting}. In contrast, fully passive models give lower masses and no unstable disks using the same observations \citep{2022SheehanVLAALMA}. To explore the dynamical state of our disks, Figure~\ref{fig:Mdisk_Mstar} compares the derived disk masses with protostellar masses from the literature (Table~\ref{table:source_prop}). When no measurement was available, we assumed $M_{*}=1$ M$_{\odot}$ (empty symbols). For close binaries for which only the combined protostar mass is reported in the literature, we assume equal mass for each component. Figure~\ref{fig:Mdisk_Mstar} shows that seven disks, including some in close binaries, approach or exceed the gravitational instability limit $M_{disk}/M_*\sim0.1$ \citep{2016KratterLodato}. In the case of IRAS 16293 B, this matches the results in \cite{2021ZamponiHotdisk} where a model of a hot and massive unstable disk was able to reproduce the 1.3-3mm fluxes, similar to the case of the Class 0/I edge-on disks L1527 IRS using observation from 0.9 mm to 7 mm  \citep{2022OhashiFormation}. Disks with $M_{disk}/M_*\gtrsim0.1$ also appear in simulations \citep{2018BateDiversity,2021LebruillyProto}, though less frequently when magnetic fields and non-ideal MHD are included \citep{2021LebruillyProto}. One might then wonder why these or many other young disks do not show for instance spiral arms features. One possibility, discussed in \cite{2022XuSpiralArmshidden}, is that the high optical depths limit the detectability of such features, as the observations cannot see all the way through the disk, plus the perturbations in density might be only of order unity. In addition, the disks might be only marginally gravitationally unstable with Toomre Q parameter $\sim$1-2 \citep{1964Toomre,2016KratterLodato}. In this case, spiral perturbations do not grow exponentially. Using our surface density and temperature profiles, we compute Q profiles assuming flat or $r^{-1.5}$ optical depth profiles, thus following the same assumptions employed for the mass constraints\footnote{the Q parameter is inversely proportional to the surface density, thus it will also change linearly with the assumed $\kappa$ which in this case is 1.} ( Figure~\ref{fig:toomre_q}). Even under conservative assumptions, several disks reach $Q\sim1-2$ or less at least in the outer parts, supporting the scenario that gravitational instabilities are playing a role during the protostellar stages. 
 

\subsubsection{Dust substructure and molecular line detection}
\label{sec:sub_mol_lines}

The only clear substructure in our sample is the ring in the Class I disk IRS 63 (Figure~\ref{fig:obs_1mm}), previously identified by \citet{2020SeguraCoxFour}, which uniquely shows $\alpha$ values above the optically thick limit well before the disk edge in our sample. Another potential substructure appears in the Class 0 disk OphA SM1 intensity profile at 3 mm, showing a deviation in the radial profile near 30 au (Figure~\ref{fig:rad_profiles}), which could be due to an annular substructure \citep{2024MaureiraSM1}. Notably, the IRS 63 ring lies beyond the optically thick region, while the SM1 deviation also occurs where $\tau_{3\rm mm}<1$ \citep{2024MaureiraSM1}. These results support the idea that annular substructures can emerge as early as the Class 0 stage but are often hidden by optically thick emission. In line with this, \cite{2025HsiehCAMPOSII_substructure} find that substructure detection sharply increases to $\sim60$\% at T$_{bol}\sim200-400$ K (Class I) for features directly visible in the 1.3 mm images, while the detection rate drops to zero for Class 0 disks. The latter show slightly elevated average brightness-to-dust temperature ratios, suggesting higher optical depths.  In Class 0 disks, large scale heights may further obscure gaps; for example, bumps in the Class 0 NGC 1333 IRAS4A profile can be explained with a gap in a hot and highly flared disk \citep{2024Guerra-AlvaradoIRAS4A1}.\\

The higher optical depths at 1.3 and 3 mm also complicates kinematic and chemical studies. Optically thick dust can cause molecular lines above the disk to appear in absorption (e.g., \citealt{2012PinedaIRAS,2018OyaSourceB,2019SahuImplications,2021GarufiALMADOT}), or remain undetected as part or all molecular emission can be masked within the optically thick dust, leading to underestimated abundances. VLA observations already demonstrate that longer wavelengths can reveal species missed in ALMA $\sim$1 mm observations \citep{2017Lopez-SepulcreIRAS4A,2020DeSimoneVLA}. Thus, abundance measurements at disk scales must account for high dust optical depths \citep{2025HsiehResolved}. Future high-sensitivity facilities such as SKA\footnote{\url{https://www.skao.int/en/science-users}} and ngVLA\footnote{\url{https://ngvla.nrao.edu/}} will be crucial to advance studies of young disk kinematics and chemistry.

\subsubsection{Slope of the temperature radial profile}

\label{sec:temp_slope}

Given that the disk emission is optically thick, the radial profile of brightness temperature can, in principle, trace the temperature radial profile and thus inform us about the dominant heating mechanism. Our analysis in Section~\ref{sec:tb_vs_td_alpha} finds that the power-law index of brightness temperature with radius spans from approximately -1.2 to 0 (Figure~\ref{fig:gamma1_sources}). These values extend both above and well below theoretical expectations for passive (-0.5) and viscous (-0.75) heating. Moreover, the slopes at 3 mm tend to be steeper than at 1.3 mm, consistent with lower optical depths at longer wavelengths, regardless of whether the disk is marginally or fully optically thick.

The variation in slope across the sample, including several disks with nearly flat profiles, may indicate significant differences in absolute optical depth. Flatter profiles likely correspond to very high optical depths. To understand this, we can consider two idealized cases under the Rayleigh-Jeans and modified blackbody approximations: optically thin emission for which $T_b\propto r^{q+p}$ with $q$ and $p$ the power-law index of the temperature and surface density as a function of radius, and optically thick emission for which $T_b\propto r^q$. It follows that when emission goes from optically thick to thin, the brightness profile becomes steeper. This is consistent with the Class I disk IRS 63, which shows the steepest slope ($\sim$-1) and has low spectral index values ($\alpha\lesssim2$) confined only to the inner disk, suggesting reduced optical depth in outer regions (Figures~\ref{fig:rad_profiles} and~\ref{fig:gamma1_sources}). However, in extremely optically thick disks, the temperature observed at each wavelength corresponds to the $\tau\sim1$ surface, which can vary in shape and altitude across wavelengths. In such cases, we may not probe the midplane at all, preventing us from probing the radial temperature profile. For instance, the edge-on Class 0 disk HH212 shows flat brightness profiles that steepen with increasing wavelength \citep{2021LinHH212}, due to the $\tau\sim1$ surface becoming less flat at longer wavelengths. Their modeling inferred a physical temperature profile with $q\sim-0.7$, despite the observed flatness. Similarly, the nearly face-on Class 0 disk IRAS 16293B shows flat profiles. \cite{2021ZamponiHotdisk} demonstrated that even in such orientation, the $\tau\sim1$ surfaces at 1.3 mm and 3 mm can lie above the midplane and remain relatively flat in the inner regions. Based on these studies, detailed modeling and multi-wavelength observations are necessary to constrain the true temperature power-law index. Still, some disks in our sample already show 3 mm slopes near -0.75 (e.g., RCrA IRS7B a, RCrA SMM1C) or -0.5 (VLA 1623B), suggesting that q may vary among disks even at similar evolutionary stages.

\subsection{Dust temperatures and iceline locations}
\label{sec:chemist_connection}

\begin{figure}
  \resizebox{\hsize}{!}{\includegraphics{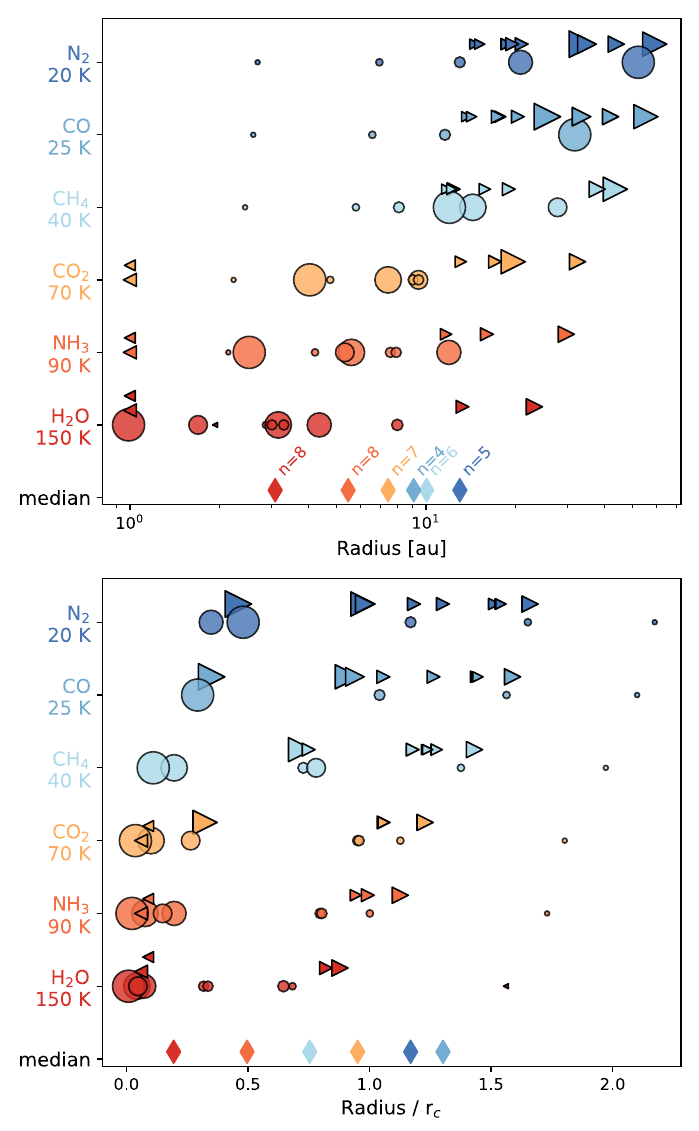}}
  \caption{Location of different icelines in the observed Class 0/I disks derived from the fit intensity profiles at 1.3 mm. Top: the location is plotted as distance from the center of the disk. Bottom panel: the location is plotted as the radius of the iceline over the characteristic disk radius r$_c$. The '$\triangleleft$'
 and '$\triangleright$' symbols correspond to upper and lower limits, respectively. The sizes of the symbols are proportional to $\log_{10}{r_c}$. The median value for each iceline (ignoring upper/lower limits) is marked with a diamond symbol. The numbers of disks considered for the median is indicated next to diamond marker at the top panel. See Section~\ref{sec:chemist_connection} for more details.}
         \label{fig:iceline_loc}
\end{figure}

Given that the high optical depths allow us to measure dust temperatures, we estimate where different icelines would be located in our Class 0/I disk sample. When doing this estimation, and based on the discussion in~\ref{sec:temp_slope}, it is important to keep in mind that the high optical depths in some disks are likely preventing us from tracing the midplane temperatures, and thus the radial location discussed here is only a first approximation. 

To obtain the location of an iceline at temperature $T_{\text{ice}}$, we use the $T^{1.3mm}_b$ profiles obtained from the model intensity profiles fit (Section~\ref{sec:profile_fit}). We find the radius $r_{\text{ice}}$ such that $T^{1.3mm}_b=T_{\text{ice}}$ provided that $\alpha\leq2$. When $\alpha>2$ the value for $r_{\text{ice}}$ is taken as a lower limit. When $T_{\text{ice}}>max(T_b)$ throughout the profile, we set $r_{\text{ice}}=1$ au as an upper limit. Figure~\ref{fig:iceline_loc} top panel shows the computed values for $r_{\text{ice}}$ for difference icelines for all the disks in the sample with 1.3 mm observations\footnote{Exceptions are IRAS 16293 A1/A2 for which the fit failed (see Table~\ref{tab:sizes_slopes_gaussian}).}. The adopted values for $T_{\text{ice}}$ should be considered approximate, as they are sensitive to variations in ice composition and local density conditions \citep{2014MartinDomenechThermalDesorp,2016FayolleN2,2019PotatovPhoto,2024VantHoffPPDChemStructure}. 

We find that the median location\footnote{The median is calculated ignoring upper/lower limits.} of the water iceline ($T_{\text{ice}}$=150 K) is $r_{\text{ice}}\sim3$ au. There is significant spread across the sample, which can be partially associated to the range of disk sizes in the sample from $\sim1$ to $\sim$ 100 au (symbol sizes). Only upper limits are found for several disks (IRS7B b, SMM1C, CXO 34, VLA 1623 Aa/Ab), as well as values as high as 8 au (L1551 IRS 5 S), and lower limits of 13 au and 23 au, for L1551 IRS5 N and IRAS 16293 B, respectively. Figure~\ref{fig:iceline_loc} bottom panel shows the  location of the icelines as a function of the ratio $r_{\text{ice}}/r_c$, in order to show what fraction of the disk size we can expect $T_d>T_{\text{ice}}$. There is a group of disks for which the water iceline is located only in the inner region of the disks ($\lesssim 0.1r_c$, IRS 63, IRS7B a and b, SMM1C, VLA1623 B and CXO 34), and another group of disks have values $\gtrsim 0.3r_c$ (IRS 7A, Elias 29, IRAS 16293 B, L1551 IRS 5 N and S, VLA1623 Aa and Ab). Despite the larger disk size fraction in the latter disks with dust temperatures above 100-150 K, emission from complex organic molecules (COMs) have only been detected for the largest hotter disks: IRAS 16293 B \citep{2005ChandlerIRAS,2012PinedaIRAS,2016JorgensenPILS}, and L1551 IRS 5 N/S \citep{2020BianchiL1551}. We remind the reader that with the exception of IRAS 16293-2422, all sources in the sample are part of the FAUST and therefore, have dedicated observations aimed at revealing COMs emission. For the compact sources RCrA IRS7A (R$_{95}\sim$7 au) and Elias 29 (R$_{95}\sim$2 au), further inspection of the ALMA spectral setups from FAUST reveals no bright, compact sources associated with COM emission at a resolution of 50 au. This could be due to the high dust optical depths, but also due to the compact disk size in these sources, as both make the detection of complex molecules more difficult. Overall, COMs have been detected in our sample only in sources for which the size of the region with dust temperature above 100 K is computed to be at least 10 au. In the case of Elias 29, \cite{2019OyaSulfur} also proposed that the lack of COMs towards this source is due to a more elevated envelope temperature ($>20$ K) during the prestellar phase which could prevent efficient depletion of the parent species of COMs (e.g., CO). Thus, the lack of COM detection in Elias 29 and IRS 7A could also be related to an intrinsic chemical difference originated in the environment \citep{2025OyaFAUSTElias}.\\ 

The other, colder icelines in Figure~\ref{fig:iceline_loc}, follow similar trends. The median $r_{\text{ice}}$ values for the  N$_2$ ($T_{\text{ice}}$=20 K) and CO ($T_{\text{ice}}$=25 K) icelines are 13 au and 9 au respectively. The location of these two icelines in most of the disks is located beyond the characteristic disk radius, which is consistent with the extensive use of C$^{18}$O for tracing keplerian rotation in young disks. 

\subsection{Disk sizes at 1.3 and 3 mm and grain growth}

The high optical depth in the observed disks prevents us from directly using the resolved spectral index to infer the dust emissivity index $\beta$, a proxy for the maximum grain sizes, throughout the disk. However, we note that the outer edges of several disks in our sample show a steep $\alpha_{1.3-3mm}$ gradient (Figure~\ref{fig:alpha_all}). This rapid increase, with $\alpha$ values changing from $\sim2$ to $\sim3$, is typically associated with the location of the fall off in the intensity profiles (Figure~\ref{fig:rad_profiles}), hence related to the transition from optically thick to  thin emission. From this, we can conclude that the optically thin dust in the outer disk region is in agreement with $\beta\approx\alpha-2\sim1$ (or higher considering non R-J effects at lower temperatures). Such values do not rule out the presence of grains that are from a few 10 $\mu$m and up to mm sizes, depending on the adopted opacity \citep{2014TestiDustEvolution,2024BirnstielReview,2024ZamponiScattering,2019CarrascoRadial}. Observations at longer wavelengths with the VLA, and in the future ngVLA/SKA, are required to overcome the high optical depth, thus allowing us to reveal the grain size distribution in these young disks. Indeed, \cite{2025RadleyVLA1623} recently derived the presence of mm-sized grains in the Class 0 and I disks in the VLA 1623 A/B and W system (Figures~\ref{fig:obs_3mm} and~\ref{fig:obs_1mm}), by combining ALMA high-resolution observations at 1.3 and 3 mm, with VLA observations at 7 mm, 1.4 cm and 3 cm. \\

An independent constraint to the grain sizes can be obtained by looking at how the disk size changes with wavelength. In this scenario, the observed sharp outer disk edge in the dust emission profile is not tracing a drop of the dust surface density, but instead a sudden change in the dust opacity, related to the presence of maximum grain sizes of about few 100 $\mu$m interior to the intensity profile fall off \citep{2019RosottiEvolutionDiscRadius}. This is because the dust opacity at a given wavelength increases sharply with maximum grain size at around $a_{max}\sim\lambda/2\pi$. Thus, the observed disk edge or 'knee' at 1.3 mm and 3 mm would be due to grain sizes larger than few 100 $\mu$m interior to the knee and smaller beyond that location. In that scenario, the disk radius is also smaller for longer wavelengths, as the sudden increase in the opacity moves with wavelength \citep{2021TazzariMulti}. In Figure~\ref{fig:size_lum}, we showed that both $R_{68}$ and $R_{95}$ are comparable to within 4-7\% for the Class 0/I disks studied here. \cite{2021TazzariMulti} finds also small differences (below 9\%) for $R_{68}$ between 0.9, 1.3 and 3 mm for Class II disks in Lupus. They concluded that the small difference could be reproduced by considering constant $\beta$ values in the range 0.5-1 over a substantial part of the disks with possible larger grains only in the very inner regions, i.e., not strong variations in the grain sizes with radius for most of the disk emitting area. According to the models presented in \cite{2021TazzariMulti} (see their Figure 8), the similarity in sizes between 1.3 and 3 mm, the high fraction of optically thick emission and the associated typical spectral index of $\sim$2 obtained for our Class 0/I disk sample can be all reproduced by also considering $\beta$ values in the range 0.5-1, similar to the conclusions for the Class II disks.

\subsection{Class 0/I Circumbinary Disks}
\label{sec:cb_disks_disc}

Figure~\ref{fig:obs_3mm} shows that circumbinary disks (CBDs) are detected only in systems with projected separations below 100 au (14–54 au in our sample), while wider multiples ($\geq$103 au) lack CBDs. This matches simulations  showing that the fraction of CBDs  with ages up to 0.1 Myr drops rapidly from about 50\% for binaries with semimajor axes $a$ in the range 10-100 au, to almost zero for $a\gtrsim100$ au \citep{2019BateStatistical,2023ElsenderCBfreq}. The same trend holds when considering other ALMA high-resolution Class 0/I surveys (eDisk; \citealt{2023OhashieDisk}, CAMPOS; \citealt{2024HsiehCAMPOSI}) as well a sample of multiples in Perseus \citep{2024ReynoldsPerseus}. Although we do not have a measurement of $a$ for the majority of these close Class 0/I multiples (see \citealt{2020MaureiraOrbits} and \citealt{2024HernandezOrbitsL1551} for exceptions), the lack of CBDs above $\sim$100 au projected separation strongly supports a real drop around the predicted threshold of $a\sim$100 au \citep{2023ElsenderCBfreq}.


 \begin{figure*}
\centering
   \includegraphics[width=14cm]{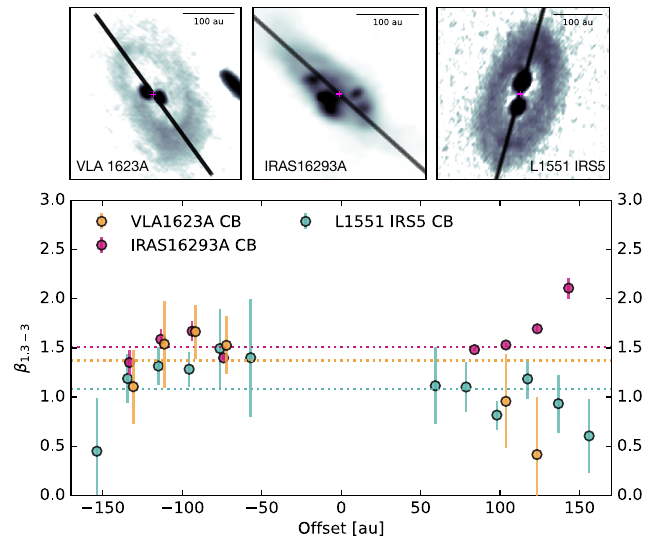}
     \caption{Derived $\beta_{1.3-3mm}$ for the CBDs in the sample. The values are calculated for a cut centered at the midpoint between the CSDs, oriented along the CBD  major axis. The cuts are displayed in the top panels. The dotted lines correspond to the mean value for each source. Positive offsets correspond to the eastern side of the cut. See Section~\ref{sec:cb_beta} for details.}
         \label{fig:cb_beta}
\end{figure*}

\subsubsection{Class 0/I CBDs optical depths and masses}
\label{sec:cb_masses}

For the three CBDs for which we have observations at 1.3 and 3 mm (VLA 1623 A, IRAS 16293 A, and L1551 IRS 5) we find that their mean spectral index are in the range 2.6-3.1 (Figure~\ref{fig:kde_alpha}). Such a difference between the spectral index in CBDs versus circumstellar disks (CSDs) implies a change in the optical depth in these populations, with lower values for CBDs. \cite{2022Maureirahotspots} studied in detail the 1.3-3 mm spectral index in IRAS 16293 A CBD. By assuming a constant value of $\beta$ throughout the CBD, the optical depth at 1.3 mm is estimated in the range $\sim$0.4-1, corresponding to 0.2-0.3 at 3 mm. This is indeed, much lower than the typical optical depths at 3 mm in the CSDs discussed above (see Figure~\ref{fig:toomre_q}), but comparable with the values observed at the CSD's edges. We can estimate the optical depth for the CBDs in the sample, by comparing the predicted temperatures due to irradiation with the observed intensities at 1.3 and 3 mm. We performed this comparison over a cut oriented along the CBD's major axis, and passing through the mid-point between the individual CSDs. Details about the intensity cuts and the temperature profile can be found in Appendix~\ref{sec:cb_appendix_tau}. The resultant optical depths are below 0.5 for L1551 IRS 5 CBD and VLA 1623 CBD at both wavelengths, while for IRAS 16293 A CBD the optical depth at 1.3 mm is higher, reaching values close to 1, in agreement with previous constraints \citep{2022Maureirahotspots}. The intensity, temperature and optical depth cut profiles are summarized in Figure~\ref{fig:cb_temp_tau}. \\

Having estimated the optical depth in these CBDs, we can measure the amount of material in young Class 0/I CBDs. Such measurements are important given the detections of circumbinary planets around binary stars \citep{2011DoyleKepler16,2018MartinCBP,2023StandingMCBP}. For VLA1623A and L1551 IRS5 CBDs, we measure the 1.3 mm integrated flux in the CBDs within a contour above 3$\sigma$, and use the optically thin approximation, adjusting for the temperature and $\beta$ based on the above discussion. Details can be found in Appendix~\ref{sec:cb_appendix_tau}. The mass underestimations due to the assumption of optically thin emission are less than 25\% for these optically thinner structures, which is less than the typical uncertainties due to the choice of dust opacity. The resultant dust masses are 94 M$_{\oplus}$ for VLA1623A CBD, and 78 M$_{\oplus}$ for L1551 IRS 5 CBD, this corresponds to $\sim0.03$ M$_{\odot}$, and $\sim0.02$ M$_{\odot}$ in gas (or $\sim$26 M$_{\text{Jup}}$), assuming a gas-to-dust ratio of 100. These masses are comparable to, or higher than the ones derived for the CSDs in these systems (Figure~\ref{fig:Mdisk_Mstar}). As shown in \cite{2022Maureirahotspots}, the circumbinary disk around IRAS 16293 A shows additional spots of bright emission surrounding the CSDs, which are consistent with localized shock heating. Taking into consideration this complex temperature structure, \cite{2022Maureirahotspots} derived a range of masses for this CBD of 0.02-0.04 M$_{\odot}$ for the gas ($\sim$20-40 M$_{\text{Jup}}$) or 67-133 M$_{\oplus}$ in solids\footnote{We adjusted the values considering the different assumed opacities in these two works.}. Overall, the gas mass range of $\sim$20-40 M$_{\text{Jup}}$ for these Class 0/I CBDs is at least comparable to that of the currently detected circumbinary planets with masses between 0.1-13 M$_{\text{Jup}}$\footnote{NASA Exoplanet Archive \url{https://exoplanet.eu/planets_binary/}}. 

\subsubsection{Dust emissivity and grain sizes in the Class 0/I CBDs}
\label{sec:cb_beta}

We use a modified black body to derive $\beta$ profiles along the major axis of the CBDs, using the optical depth and temperature profiles derived in the previous section and the full Planck function. The results are shown in Figure~\ref{fig:cb_beta}, where we binned the $\beta$ profiles in bins of 20 au. The mean $\beta_{\text{CBD}}$ for VLA 1623 A and IRAS 16293 A (Class 0) are $1.37\pm0.15$ and $1.51\pm0.01$, respectively. For these binaries, the derived values can be taken as a conservative lower limit given that the internal luminosities of the binaries therein also include a third protostellar source nearby (see Table~\ref{table:source_prop}). For the Class I L1551 IRS5, the mean $\beta_{\text{CBD}}$ is $1.1\pm0.07$. The results for the CBDs suggests a possible trend of higher $\beta$ values for the Class 0 sources. The measurements of $\beta$ for the Class 0 CBDs suggests either ISM-like grains or grains that have grown up to few 100 $\mu$m, as similar $\beta$ are expected in both scenarios \citep{2019AgurtoGangasPeremb50,2010RicciDust}. On the other hand, the Class I circumbinary disk L1551 IRS 5 could host larger mm-sized grains, based on the estimated lower $\beta$ values. This is in agreement with recent findings of low $\beta\lesssim$1 values in the dusty and wide outflow cavity of this source, extending up to 2,000 au  \citep{2025SabatiniL1551irs5Cavitydust}. These grains might then have been lifted from the circumbinary disk. Finally, we note that the $\beta_{\text{CBD}}$ for VLA 1623 A and L1551 IRS5 are also in agreement with the values found for the envelope around these sources corresponding to $\beta_{\text{env}}\sim1.43\pm0.05$ and $\beta_{\text{env}}\sim0.94\pm0.04$, respectively \citep{2025CacciapuotiFAUSTbeta}. 







\section{Summary and Conclusions}
\label{sec:conclusions}

We present and analyze 1.3 and 3 mm ALMA continuum observations with a median resolution of 7.5 au toward a sample of 16 Class 0 and I disks in single and multiple systems, and across several molecular clouds.  The sample is more representative of the early protostellar stages, with 14/16 sources having bolometric temperatures $<$ 200 K. In addition, about a third of the disks analyzed in this sample (Elias 29, L1551 IRS 5 N/S and IRAS 16293 A1/A2) have bolometric luminosities $>$ 20 L${\odot}$. Such luminosities are higher than those typically derived for Class II pre-main sequence stars, but they may be indicative of protostars undergoing higher accretion rates, as predicted for early stages \citep{2014DunhamPPVI} and confirmed by direct measurements in some Class I protostars (e.g., \citealt{2023FiorellinoMassAccrClassI}). The observations reveal individual disks with a variety of sizes ranging from $R_{95}\lesssim3$ au to $100$ au. The observations also reveal four circumbinary disk (CBD) structures reaching $\sim$100 au scales. The observations allowed us to built resolved spectral index maps to investigate the optical depth in the disk structures. Our results and conclusions are summarized as follows:\\

\noindent\emph{Class 0/I disks:} We find that the disks in our sample are mostly optically thick at 1.3 and 3 mm. The elevated optical depth results in most of the estimated disk masses in the range 30-900  M$_{\oplus}$, corresponding to gas\footnote{Assuming gas-to-dust ratio of 100} masses in the range 0.01-0.3 M$_{\odot}$. Such values allow for a fraction of the disks to be marginally gravitationally unstable, with derived Toomre Q values $\lesssim2$ for the outer disk regions. Some disks appear to have temperatures that are in excess of what is expected from irradiation alone. The median location for the water iceline ($T_{\text{ice}}=150$) K is $\sim3$ au, with lower limits for the hotter disks from 10 to 20 au. This suggests the need to account for other types of heating such as accretion/viscous heating for the hotter disks. \\

\noindent\emph{Comparison with Class II disks:} Class 0/I disks exhibit low spectral indices ($\alpha_{CSD}\sim2$), similar to Class II disks, and follow the same trend of increasing $\alpha$ with disk size. As class II disks, Class 0/I disk sizes at 1.3 and 3 mm differ by less than 10\%, indicating minimal wavelength-dependent structure in this wavelength range. A clear correlation between disk size and millimeter luminosity is observed, with a slope comparable to that of Class II disks. However, Class 0/I disks are 10$\times$ brighter at a given size, consistent with their higher brightness temperatures. This luminosity offset can be explained with a higher fraction ($\sim$1) of optically thick emission for the younger disks, though some cases may also require higher temperatures. Consequently, assuming similar dust properties, Class 0/I disks likely host higher dust masses than their more evolved Class II counterparts.\\

\noindent\emph{Class 0/I circumbinary disks:} CBDs are detected only around binaries with projected separations $\leq100$ au, consistent with other observations from the literature and numerical simulations indicating a sharp decline in the occurrence of CBD for larger separations. These CBDs typically exhibit optical depths $\lesssim1$ at 1.3 and 3 mm, which accounts for their higher spectral indices ($\alpha_{CBD}\sim3$) compared to individual disks. Estimated CBD masses range from 80 to 130 M$_{\text{Earth}}$, comparable to or exceeding those of known circumbinary planets. CBDs dust emissivity indices $\beta$ derived from 1.3 and 3 mm data fall between 1–1.5, with the Class I CBD L1551 IRS 5 showing $\beta\sim$1, suggesting larger, mm-sized grains, while the two Class 0 CBDs (VLA1623 and IRAS 16293 A) exhibit slightly higher $\beta$ values ($\sim1.4-1.5$), which can be explained with ISM-like grain populations. \\

The derived high optical depths and high optical depth fractions at both 1.3 and 3 mm toward the sample of individual Class 0/I disks strongly support the possibility that annular substructures related to the planet formation process may emerge as early as the Class 0 stage, but remain difficult to detect with current observations (e.g., \citealt{2024MaureiraSM1,2025HsiehResolved}). Optically thick dust also hinders the estimation of grain sizes within these disks and poses challenges for studying the disk kinematics and chemistry at the location of the optically thick dust. In this study, we could only obtain an estimation of $\beta^{\text{edge}}_{\text{disk}}\gtrsim1$ at the disk edge, based on $\alpha^{\text{edge}}_{\text{disk}}\gtrsim3$ therein. Observations at longer wavelengths are necessary to overcome these issues (e.g., \citealt{2020DeSimoneVLA,2022OhashiFormation,2025RadleyVLA1623}) and thus future observations with SKAO and ngVLA along with more sensitive observations with ALMA to reach wider and fainter populations, will be key for advancing our understanding of early disk and planet formation and evolution. 

\begin{acknowledgements}

MJM is grateful to Munan Gong, Kedron Silsbee, Matthew Bate, and Til Birnstiel for valuable conversations that contributed to the improvement of this work.

This work was supported by the Max-Planck Society.
D.J. is supported by NRC Canada and by an NSERC Discovery Grant. 
NC acknowledges support from the European Research Council (ERC) under the European Union Horizon Europe programme (grant agreement No. 101042275, project Stellar-MADE).
LP, EB, GS, and ClCo acknowledge the PRIN-MUR 2020  BEYOND-2p (Astrochemistry beyond the second period elements, Prot. 2020AFB3FX), the project ASI-Astrobiologia 2023 MIGLIORA (Modeling Chemical Complexity, F83C23000800005), the INAF-GO 2023 fundings PROTO-SKA (Exploiting ALMA data to study planet forming disks: preparing the advent of SKA, C13C23000770005), the INAF Mini-Grant 2022 “Chemical Origins” (PI: L. Podio), the INAF Mini-Grant 2023 TRIESTE (``TRacing the chemIcal hEritage of our originS: from proTostars to planEts''; PI: G. Sabatini) and the National Recovery and Resilience Plan (NRRP), Mission 4, Component 2, Investment 1.1, Call for tender No. 104 published on 2.2.2022 by the Italian Ministry of University and Research (MUR), funded by the European Union – NextGenerationEU– Project Title 2022JC2Y93 Chemical Origins: linking the fossil composition of the Solar System with the chemistry of protoplanetary disks – CUP J53D23001600006 - Grant Assignment Decree No. 962 adopted on 30.06.2023 by the Italian Ministry of Ministry of University and Research (MUR). EB aknowledges contribution of the Next Generation EU funds within the National Recovery and Resilience Plan (PNRR), Mission 4 - Education and Research, Component 2 - From Research to Business (M4C2), Investment Line 3.1 - Strengthening and creation of Research Infrastructures, Project IR0000034 – “STILES - Strengthening the Italian Leadership in ELT and SKA”.

\end{acknowledgements}

%
%

\bibliographystyle{aa} 
\bibliography{bibliography.bib} 

\begin{appendix}

\section{Self-calibration and imaging}
\label{sec:self-calibration-app}

\begin{table*}
\caption{Continuum images properties}
\label{table:maps_properties}      
 \centering
\begin{tabular}{l c c c r r r r r }
\hline\hline
Target Field & Band & robust & taper &rms  &bmaj & bmin & bpa &other sources  \\  
& & & (mas$\times$mas, deg.)& ($\mu$mJy beam$^{-1}$) &(")&(")&(deg.)&in the field\\ 
\hline                        %
Elias 29	&	6	&	0	&	15$\times$10,0	&	13.70	&	0.040	&	0.035	&	57.4	&	-	\\
	&	3	&	0	&	30$\times$30,0	&	5.83	&	0.050	&	0.043	&	-63.4	&	-	\\
L1551 IRS5 N/S	&	6	&	-0.5	&	-	&	77.10	&	0.046	&	0.033	&	-15.4	&	-	\\
	&	3	&	0	&	-	&	21.50	&	0.067	&	0.043	&	-32.5	&	-	\\
RCrA IRS7B a,b	&	6	&	0	&	-	&	35.00	&	0.061	&	0.042	&	53.2	&	\small{SMM1C, CXO 34, IRS7A}\\
	&	3	&	0	&	35$\times$35,0	&	9.88	&	0.056	&	0.052	&	16.6	&	\small{SMM1C, CXO 34, IRS7A }\\
   &&&&&&&& \small{CrA24, RCrA SMM2}	\\
VLA 1623 A/B	&	6	&	0	&	-	&	16.60	&	0.050	&	0.036	&	80.7	&	\small{VLA 1623 W}	\\
	&	3	&	0	&	-	&	9.25	&	0.060	&	0.042	&	-79.5	&	\small{VLA 1623 W, SM1}	\\
IRS 63	&	6	&	0	&	-	&	13.80	&	0.043	&	0.037	&	86.0	&	-	\\
	&	3	&	0	&	-	&	7.13	&	0.047	&	0.037	&	-88.1	&	-	\\
IRAS 16293-2422 A/B	&	6	&	0	&		&	104.00	&	0.114	&	0.069	&	-88.2	&	-	\\
	&	3	&	0.5	&	-	&	15.00	&	0.048	&	0.046	&	79.3	&	-	\\
\hline
\end{tabular}

\tablefoot{The columns bmaj, bmin, and bpa correspond to the synthesized beam major axis, minor axis and position angle, respectively. The rms corresponds to the rms at the phase center. }
\end{table*}

\begin{table*}
\caption{Continuum images for spectral index properties}
\label{table:maps_properties_foralpha}      
 \centering
\begin{tabular}{l c c c r r r r r }
\hline\hline
 Target field\tablefoottext{a}	&	bmaj	&	bmin	&	bpa	&	band 6 rms 	&	band 3 rms\\
&(")&(") &(deg.) &  ($\mu$mJy beam$^{-1}$)  & ($\mu$mJy beam$^{-1}$) \\

\hline                        %

Elias 29	&	0.045	&	0.037	&	-60	&	13.90	&	6.35	\\
L1551 IRS5 N/S	&	0.067	&	0.045	&	-30	&	66.70	&	21.50	\\
RCrA IRS7B a,b	&	0.063	&	0.052	&	54	&	36.70	&	9.86	\\
VLA 1623 A/B	&	0.06	&	0.043	&	-80	&	17.50	&	9.75	\\
IRS 63	&	0.048	&	0.039	&	80	&	13.90	&	7.17	\\
IRAS 16293-2422 A/B	&	0.114	&	0.069	&	-88	&	104.00	&	24.00	\\
\hline
\end{tabular}

\tablefoot{The columns bmaj, bmin, and bpa correspond to the synthesized beam major axis, minor axis and position angle, respectively. The rms corresponds to the rms at the phase center. 
\tablefoottext{a}{See Table~\ref{table:maps_properties} for additional sources in the same field.}
}

\end{table*}

We self-calibrated the long-baselines observations iteratively using CASA and the taks tclean and gaincal. We use the 'multiscale' deconvolver with 3 or 4 scales depending on the source for the first cleaning (pre-selfcal) as well as the ones performed after each phase-only self-calibration solution is applied. For both phase and amplitude self-calibration, we image with a robust parameter of 0.5. We started with phase-only self-calibration, performing seven iterations with decreasing solint intervals from 'inf' (with combine$=$scan) down to 'int'. Then, for each source and band, we kept the solution with the shortest solint interval that would still result in improved S/N and visible reduction of artifacts. In most cases this interval was 30s, but in somes cases smaller solint still resulted in improvements (Elias 29 and VLA 1623 both in band 6). After applying these solutions, we performed another round of cleaning using the 'mtmfs' deconvolver with nterms$=2$ which generally resulted in further improved image fidelity. After this process of phase-only self-calibration the S/N improved from 10\% up to 40\% in the band 3, and from 15\% up to a factor of 3.5 in the band 6. We then continued with amplitude self-calibration also using  the'mtmfs' deconvolver with nterms$=2$. We performed two steps with solint values of $inf$ and 600s. In each step the phase solution was updated using the shortest successful solint value from the phase-only process. We kept the amp+phase selfcal solution only if the S/N remained the same or improved, there were no artifacts or changes to the flux scale (with no other apparent change in the image). Only few  observations met this criterion. For all of them, the solution with solint$=$inf was used and the improvement in S/N over the phase-only solution was only marginal (1-5\%).\\

We then self-calibrated the long baselines and FAUST compact configuration continuum observations together. The self-calibration for the FAUST observations consisted of rounds of solint intervals with 'int' for phase-only solutions, followed by amplitude+phase solutions computed per-scan. For the combined self-calibration, we averaged both datasets in time and frequency. We averaged in frequency up to the maximum that would not result in a reduction of the intensity larger than $\sim$1\% for sources away from the phase center\footnote{\url{https://safe.nrao.edu/wiki/pub/Main/RadioTutorial/BandwidthSmearing.pdf}}. The final channel widths was 30 MHz or 60 MHz, depending on where additional sources appeared in the field of view. Due to a PI error in the input of the coordinates for the long-baselines observations, for all targets except IRS 63 there was an offset between the phase center of the compact configuration and the extended configuration. The shift varied from 0.2" up to 6". To account for this, we used the 'mosaic' gridder option in tclean with mosweight$=$False in all the imaging (including self-calibration) for these targets. Similar to the long-baselines only procedure, we used 3 or 4 scales for cleaning. We explored robust and uv-taper values to match the resolution we obtained from the extended configuration observations. This resulted in typical robust values of 0. We performed six successive steps of phase-only self-calibration. For this first round, we used the final image after self-calibration of the extended configuration as the model with 'inf' as the solint interval. This aligned the two datasets to the peak position of the extended configuration and thus corrects for differences in the peak positions across datasets arising from proper motion or astrometric errors of the observations. The resultant model was then used for the next steps reducing the solint intervals from 60s to 9s, always keeping the last solution that resulted in improved S/N and image fidelity. After these solutions are applied we performed another round of cleaning using the 'mtmfs' deconvolver with nterms$=$2. After this process of phase-only self-calibration the S/N improved only marginally (few percent) in some cases and up to 20\% in others. The improvements were similar in both bands. We then continued with amplitude self-calibration also using the 'mtmfs' deconvolver with nterms$=2$. We first explored three consecutive steps for amplitude self-calibration. In all of them we use solint$=inf$. In the first and second one we combine all spectral windows and use solnorm$=$True and False, respectively. For the third and final we keep solnorm$=$False and attempt a solution per spectral window. By making solnorm$=$True in the first round we are normalizing the solution to an average to correct for the temporal fluctuations without changing the flux scale. In half of the cases, this process failed at the first step, with very obvious artifacts in the resultant image. For the other half, we kept the solutions of the third step. For the ones that failed, we tried amplitude self-calibration using only the two last steps (i.e., keeping solnorm$=$False), which produced acceptable results for two more cases for which we kept the solutions per spectral window. For the rest, amplitude was not applied as the resultant images always showed significant new artifacts, reduced S/N or changes in the overall flux scale. For all the cases in which amplitude self-calibration was applied the S/N remained the same or improved from a few up to 10\%. In some cases (e.g., RCrA IRS7B field band 6) the final image has very weak background emission which contributes to the noise but that, after comparing with the compact configuration only data, it correlates with real extended emission. This was also observed in some fields of the ALMA-IMF Large Program \citep{2022GinsburgIMF}. We extended the mask in the cases in which this extended emission was fairly simple and present at the source position and performed a final clean. Table~\ref{table:maps_properties} summarizes the final rms and beam sizes of the final images.

\section{RCrA IRS 7A non-dust emission at 3 mm}
\label{sec:irs7a_nondust}

The continuum emission towards RCrA IRS 7A at 3 mm in this study shows two elongated jet-like features to the south-east (SE) and west (W) of the protostar (Figure~\ref{fig:obs_3mm}). The peak intensity of the SE feature is  86.7 $\mu$mJy beam$^{-1}$ ($\sim5.7$ K), while for the W feature is 182 $\mu$mJy beam$^{-1}$ ($\sim9.8$ K) measured in the image using CARTA \citep{2021ComrieCARTA}. Such features are not detected at 1.3 mm (Figure~\ref{fig:obs_1mm}). Using the $3\sigma$ value for the undetected 1.3 mm intensity ($\sigma\approx52$ $\mu$mJy beam$^{-1}$), the upper limit for the 1.3 - 3mm spectral index for the SE and W features are $\alpha\lesssim0.7$ and $\alpha\lesssim-0.2$, respectively. Such a low and even negative spectral index is in agreement with free-free emission from a radio jet \citep{2018AngladaRadioJetReview}. 

We also inspected the fluxes measured with the most extended configuration of the FAUST observations ($\sim$0.3" or 45 au). The compact source in those observation shows some tails of extended emission in the SE and W directions at both wavelenghts \citep{2024SabatiniOutflowFAUST}. The derived spectral index from the integrated flux in the SE and W features therein are $\alpha\sim0.3$ and $\alpha\sim1.6$, respectively. While the values for the SE feature are low and in agreement with free-free emission, higher value in the lower resolution observations towards the W feature are possibly due to more extended emission from dust emission being detected in this more compact configuration, which becomes weaker in the higher-resolution observations.  \\

Towards the disk, we derive a mean spectral index of $\sim$0.6, which also supports significant contribution from non-dust emission in the disk (Figure~\ref{fig:alpha_all}), in agreement with previous 3.0-3.7 cm observations with the VLA \citep{2014LiuVLAMonitoring} as well as the detection in the VLA Sky Survey (VLASS) at 2-4 GHz with a resolution of 2-3" \citep{2020LacyVLASS}.


\section{Disk center, position angle and inclinations}

Table~\ref{table:source_positions} summarizes the coordinates for the center of the disk from a 2D Gaussian fit to the 3 mm emission. The respective position angles and inclination derived assuming circular geometry.

\begin{table*}
\caption{Disk center, position angle and inclination from 2D Gaussian fit}
\label{table:source_positions}      
 \centering
\begin{tabular}{l r r c c c }
\hline\hline
Source & R.A. & Decl. & PA & inclination \\    
 & (J2000)  & (J2000) & (deg.) & (deg.)  \\  
\hline                        %
RCrA IRS7B a	&	19:01:56.420	&	-36:57:28.663	&	115.5	$\pm$	0.23	&	67.1 $\pm$ 0.2	\\
RCrA IRS7B b	&	19:01:56.385	&	-36:57:28.113	&	116	$\pm$	1.1	&	65.7	$\pm$ 1.2 \\
RCrA SMM1C 	&	19:01:55.305	&	-36:57:17.306	&	174.5	$\pm$	0.4	&	78.1 $\pm$ 0.5	\\
RCrA IRS7A 	&	19:01:55.331	&	-36:57:22.688	&	66.1	$\pm$	3.5	&	56.1	$\pm$ 2.6\\
CrA 24	&	19:01:55.619	&	-36:56:51.770	&	15	$\pm$	1.7	&	72.7	$\pm$ 2.2 \\
CXO 34	&	19:01:55.787	&	-36:57:28.323	&	58.6	$\pm$	0.6	&	64.6	$\pm$ 0.6\\
Elias 29	&	16:27:09.410	&	-24:37:19.369	&	25.4	$\pm$	6.3	&	28.7 $\pm$ 2.9	\\
VLA 1623 B	&	16:26:26.303	&	-24:24:30.827	&	42	$\pm$	0.1	&	76.0	$\pm$ 0.1 \\
VLA 1623 W	&	16:26:25.629	&	-24:24:29.708	&	10.1	$\pm$	0.2	&	81.7 $\pm$ 0.2	\\
SM1	&	16:26:27.852	&	-24:23:59.723	&	147	$\pm$	35	&	15.6	$\pm$ 9.2\\
IRAS 16293 B	&	16:32:22.611	&	-24:28:32.601	&	173.4	$\pm$	8	&	26.2$\pm$ 3.5	\\
IRS 63	&	16:31:35.654	&	-24:01:30.058	&	148.2	$\pm$	1.0	&	43.2 $\pm$ 0.6	\\
\hline                        %
RCrA SMM2 a	&	19:01:58.555	&	-36:57:09.212	&	point source	&	...	\\
RCrA SMM2 b	&	19:01:58.560	&	-36:57:09.139	&	point source&	...	\\				
\hline
VLA 1623 A Aa	&	16:26:26.395	&	-24:24:30.922	&	30.9	$\pm$	0.6	&	58.8 $\pm$ 0.5		\\
VLA 1623 A Ab	&	16:26:26.380	&	-24:24:30.995	&	31.5	$\pm$	0.6	&	59.0$\pm$ 0.5		\\
\hline                          %
IRAS 16293 A A1	&	16:32:22.878	&	-24:28:36.684	&	119.8	$\pm$	3.6	&	58.6 $\pm$ 3.5	\\
IRAS 16293 A A2	&	16:32:22.851	&	-24:28:36.647	&	137.9	$\pm$	1.0	&	74.3$\pm$ 1.4	\\
\hline                        %
L1551 IRS5 S	&	04:31:34.168	&	18:08:04.272	&	143.8	$\pm$	1.1	&	39.7	$\pm$ 0.7\\
L1551 IRS5 N	&	04:31:34.164	&	18:08:04.636	&	148.7	$\pm$	0.5	&	44.1	$\pm$ 0.2\\
\hline                        %

\end{tabular}

\tablefoot{
   The reported values come from the fit to the 3 mm emission for all sources except RCrA IRS 7 A and Elias 29. For the latter the values correspond to the 1.3 mm fit as the 3 mm shows evidence of significant free-free contamination. Inclination is derived assuming a circular geometry for the disk. \\
    }

\end{table*}

\section{Alignment between 1.3 mm and 3 mm}
\label{ap:alignemnt}

\begin{figure*}
   \centering
     \includegraphics[width=0.7\textwidth]{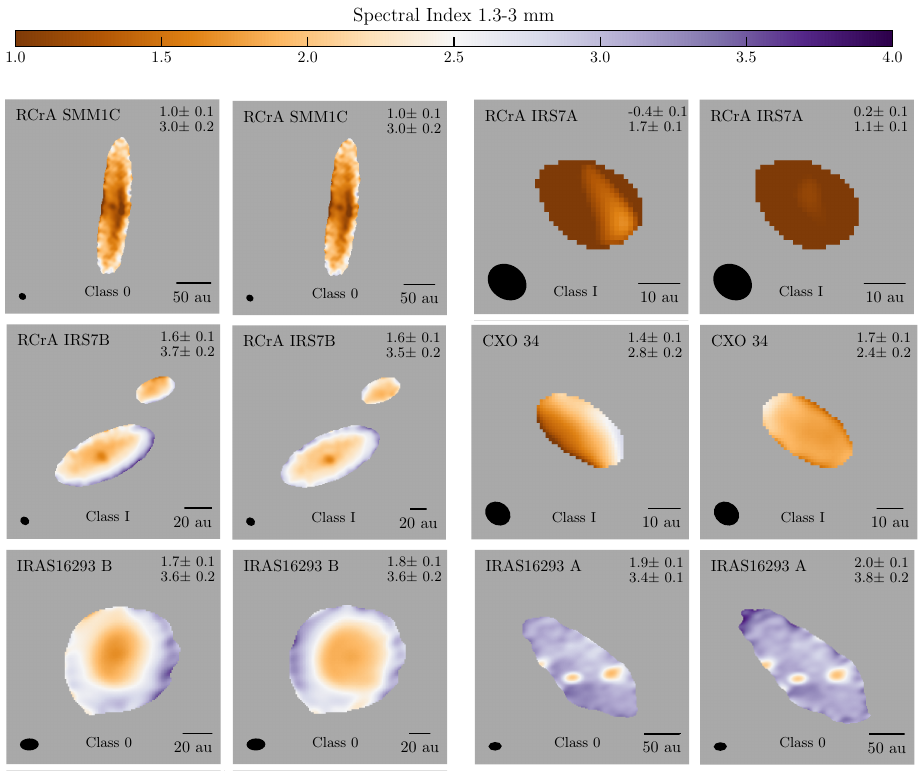}
      \caption{Spectral Index maps before and after applying a shift to the 1.3 mm image to correct for systematic gradients observed in the fields of Corona Australis and IRAS 16293 A/B (Ophiuchus). The left and right images corresponds to before and after correction, respectively. The upper right corner contains the minimum and maximum values of the spectral index in the map, considering only the pixels with errors up to 0.2.}
         \label{fig:alpha_befor_after_align}
\end{figure*}

We use the cross-correlation between the 1.3 mm and 3 mm images in order to correct for small offsets which led to visible gradients in the spectral index maps for sources in the field of Corona Australis and IRAS 16293 A/B in Ophiuchus. The cross-correlation was done individually for the sources in these fields to account for both systematic shifts in the field as well as shifts due to individual astrometric motions. Table~\ref{table:shifts} summarizes the resultant shifts of the 1.3 mm image with respect to the 3 mm image. Figure~\ref{fig:alpha_befor_after_align} shows the spectral index before and after applying the corresponding shifts. The most visually obvious gradients (IRS7B, IRS7A and CXO 24) are removed after the alignment. In the case of IRAS 16293 B some asymmetry remains, which is expected given that the source has an azimuthal asymmetry at 1.3 mm and 3 mm (Figures~\ref{fig:obs_1mm} and~\ref{fig:obs_3mm}). RCrA SMM1C shows no major change, while the circumbinary emission towards IRAS 16293 A shows a smoother distribution overall. We note that none of the results in this work are significantly affected by this small correction. 

\begin{table}
\caption{Shifts applied to the 1.3 mm image with respect to the 3 mm image to correct misalignemnts}
\label{table:shifts}      
 \centering
\begin{tabular}{l c c }
\hline\hline
Source & Offset R.A & Offset Dec.  \\    
 & (mas)  & (mas)   \\  
\hline                        %
RCrA IRS7B a,b & 7.5 & 7.5 \\
RCrA SMM1C & 0 & -7.5 \\
CXO 34 & 15 & 0 \\
RCrA IRS7A & 15 & 0 \\
\hline 
IRAS 16293 B & 20 & 10 \\
IRAS 16293 A & 0 & 10 \\

\hline                        %

\end{tabular}

\end{table}

\section{Spectral index error maps}

\begin{figure*}[t!]
   \centering
     \includegraphics[width=0.8\textwidth]{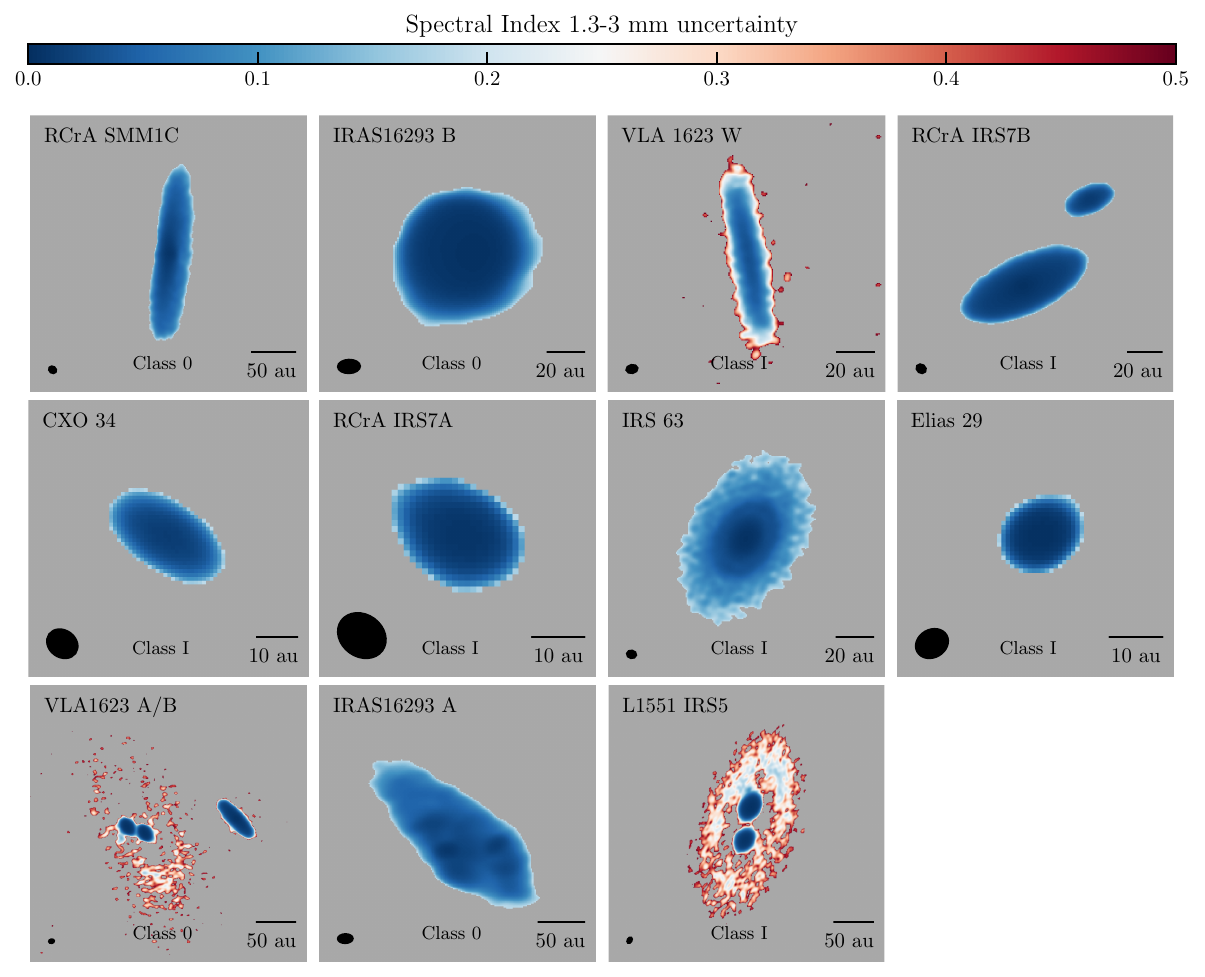}
      \caption{1.3-3 mm spectral index statistical uncertainty maps corresponding to the spectral index maps in Figure~\ref{fig:alpha_all}.}
         \label{fig:alpha_error_all}
\end{figure*}

Figure~\ref{fig:alpha_error_all} shows maps of the statistical uncertainty of the 1.3-3 mm spectral index maps in Figure~\ref{fig:alpha_all}. 

\section{Fit to intensity profiles of extended disks}

\begin{figure*}
   \centering
     \includegraphics[width=1\textwidth]{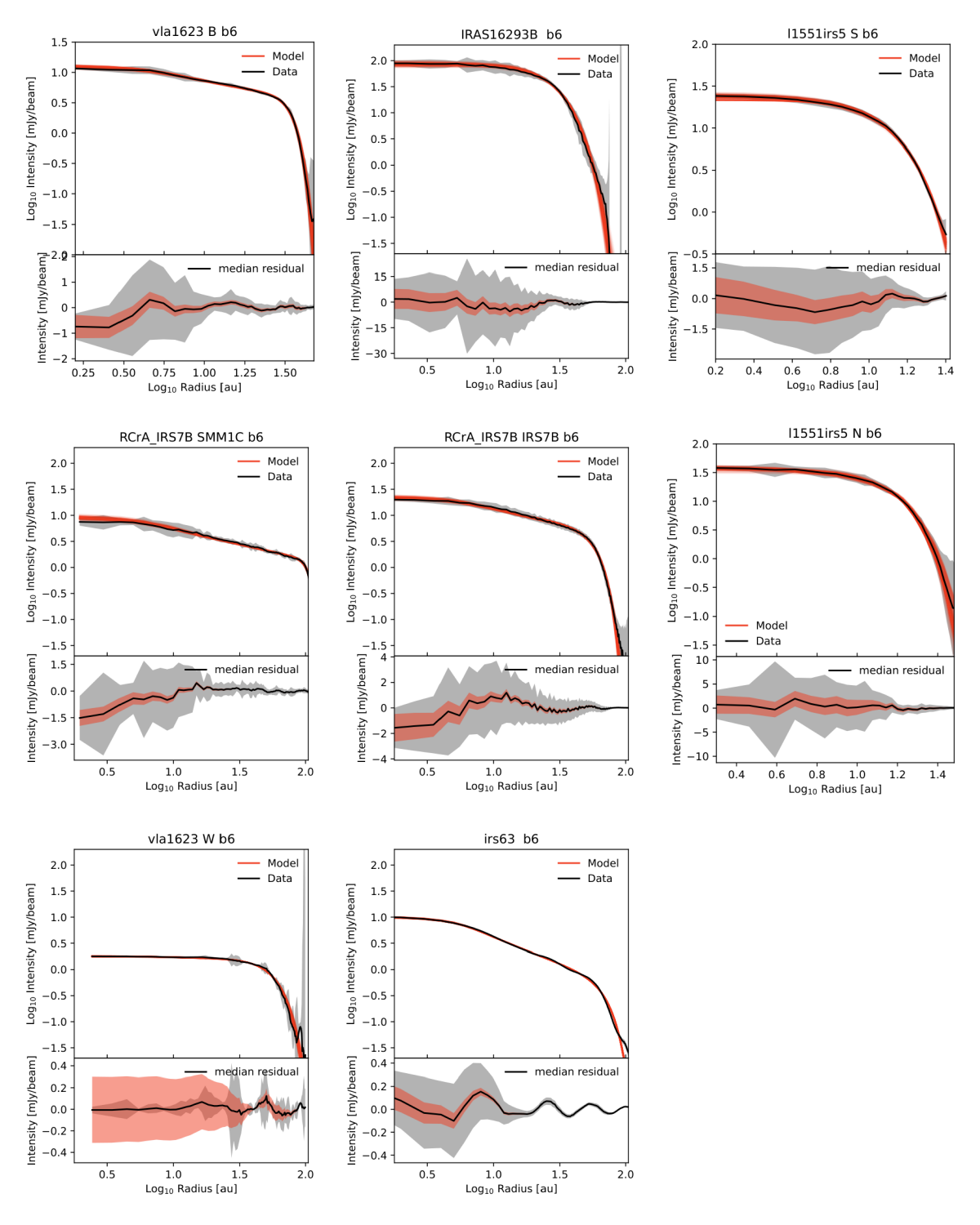}
      \caption{Fits to the 1.3 mm intensity profiles and residuals. The orange curves at the top of each panel show 100 profiles drawn randomly from the posterior distributions. The observed profile is shown with a black solid line. The resultant median residual is shown at the bottom of each panel (solid black line) as well as 1$\sigma$ dispersion around the median residual (orange shaded area).  The gray shaded area corresponds to the $1\sigma$ uncertainty of the intensity profile in all panels.}
         \label{fig:model_obs_res_plots_b6}
\end{figure*}

\begin{figure*}
   \centering
     \includegraphics[width=1\textwidth]{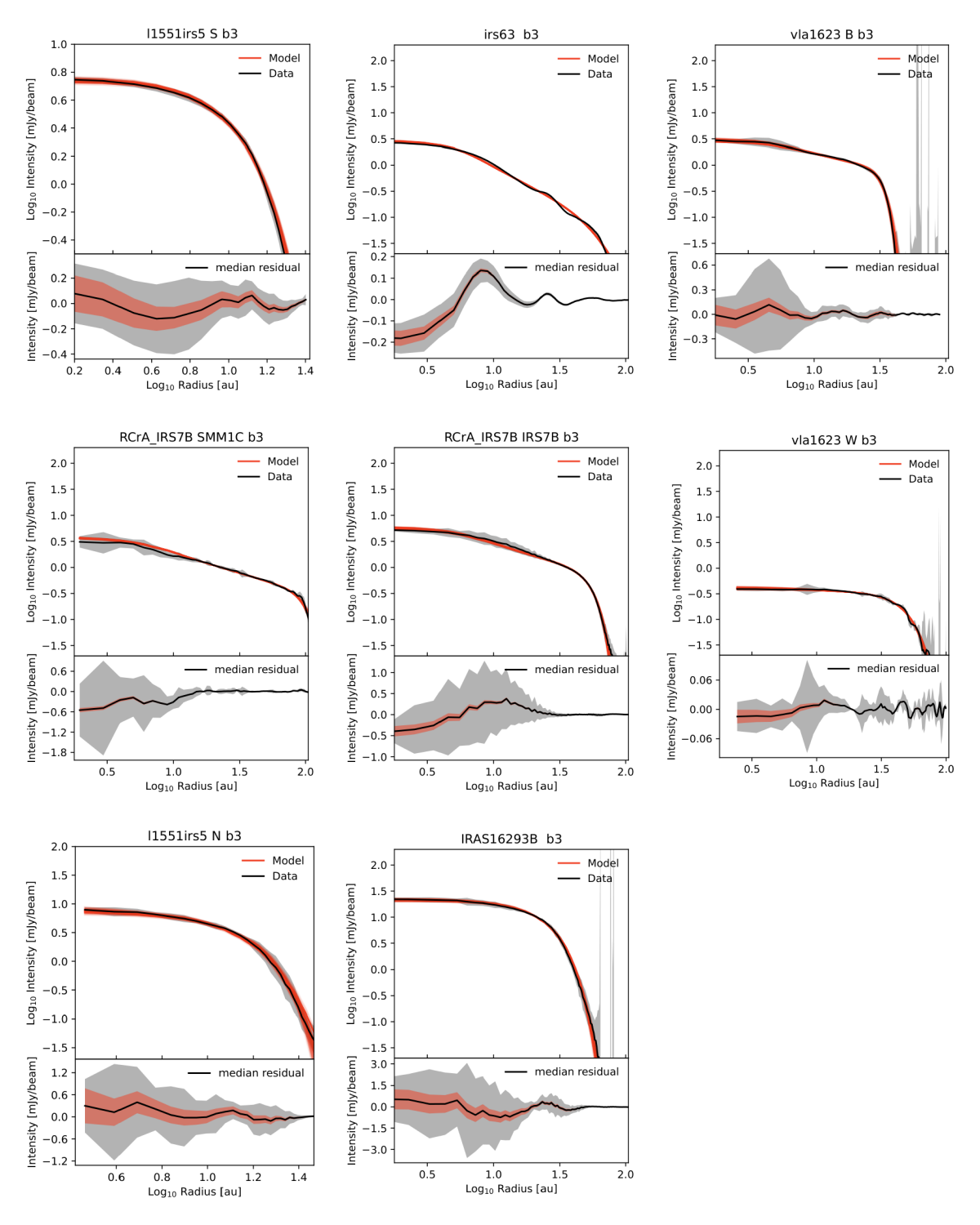}
      \caption{Same as Figure~\ref{fig:model_obs_res_plots_b6} for the 3 mm intensity profiles.}
         \label{fig:model_obs_res_plots_b3}
\end{figure*}

Figures~\ref{fig:model_obs_res_plots_b6} and~\ref{fig:model_obs_res_plots_b3} show the fit to the intensity profiles and residuals for the observations at 1.3 mm and 3 mm, respectively.

\begin{table*}
   \renewcommand*{\arraystretch}{1.4}
   \caption{Disk properties derived from fit to intensity profile using the modified self-similar profile}
   \label{tab:sizes_slopes}
   \centering
   \begin{tabular}{llcccccc}
   \hline\hline
   Source & $\lambda$ & $F_{\rm tot}$\tablefootmark{a}  & r$_c$ & $\gamma_1$ & $\gamma_2$ & R$_{95}$ & R$_{68}$\\ 
   & (mm) & (mJy) & (au) &  &   & (au) &(au)  \\ 

   \hline
   RCrA IRS7B a &1.3 & $830.21^{+10.06}_{-9.60}$ &  $60.4^{+0.8}_{-0.8}$ & $-0.52^{+0.02}_{-0.02}$ & $4.59^{+0.26}_{-0.24}$ &$65.0^{+0.5}_{-0.5}$ & $44.5^{+0.5}_{-0.5}$\\
   &3 & $135.94^{+1.27}_{-1.23}$ &  $57.2^{+0.4}_{-0.4}$ & $-0.70^{+0.01}_{-0.01}$ & $5.07^{+0.25}_{-0.26}$ &$59.5^{+0.5}_{-0.5}$ & $40.0^{+0.5}_{-0.5}$\\
   RCrA SMM1C&1.3 & $795.47^{+17.92}_{-14.69}$ &  $109.1^{+1.8}_{-1.3}$ & $-0.55^{+0.02}_{-0.02}$ & $10.31^{+1.98}_{-1.78}$ &$106.5^{+2.5}_{-2.0}$ & $80.0^{+1.0}_{-1.0}$\\
   &3 & $193.19^{+3.52}_{-2.86}$ &  $105.4^{+1.5}_{-1.3}$ & $-0.73^{+0.01}_{-0.01}$ & $5.87^{+0.55}_{-0.51}$ &$107.0^{+2.5}_{-2.5}$ & $73.5^{+1.0}_{-1.0}$\\
    VLA 1623 B  & 1.3 & $336.06^{+4.19}_{-4.20}$ &  $35.7^{+0.3}_{-0.3}$ & $-0.47^{+0.02}_{-0.02}$ & $8.88^{+1.08}_{-0.92}$ &$35.0^{+0.5}_{-0.5}$ & $26.5^{+0.5}_{-0.5}$\\ 
 &    3 & $68.00^{+1.01}_{-0.96}$ &  $33.7^{+0.3}_{-0.4}$ & $-0.54^{+0.03}_{-0.03}$ & $8.29^{+1.52}_{-1.18}$ &$33.0^{+0.5}_{-0.5}$ & $24.5^{+0.5}_{-0.5}$\\ 
  VLA 1623 W  & 1.3 & $293.09^{+4.21}_{-3.73}$ &  $59.8^{+0.5}_{-0.5}$ & $-0.05^{+0.01}_{-0.01}$ & $3.50^{+0.15}_{-0.14}$ &$73.5^{+1.0}_{-1.5}$ & $50.0^{+0.5}_{-0.5}$\\ 
  &   3 & $43.71^{+0.40}_{-0.39}$ &  $50.3^{+0.7}_{-0.6}$ & $-0.08^{+0.02}_{-0.02}$ & $3.1^{+0.12}_{-0.12}$ &$64.5^{+1.0}_{-0.5}$ & $43.5^{+0.5}_{-0.5}$\\ 
  Oph A SM1 &   3 & $19.11^{+0.51}_{-0.43}$ &   $31.1^{+1.3}_{-1.5}$ & $-0.63^{+0.04}_{-0.03}$ & $2.27^{+0.29}_{-0.26}$ & $43.5^{+2.5}_{-2.5}$& $25.0^{+1.0}_{-0.5}$\\
 IRAS 16293 B  &  1.3 & $1113.39^{+16.93}_{-16.69}$ &  $27.2^{+2.2}_{-2.6}$ & $-0.12^{+0.12}_{-0.09}$ & $2.27^{+0.25}_{-0.24}$ &$41.5^{+0.5}_{-0.5}$ & $26.0^{+0.5}_{-0.5}$\\ 
  &   3 & $209.36^{+2.76}_{-2.73}$ &  $25.5^{+1.6}_{-1.9}$ & $-0.15^{+0.10}_{-0.09}$ & $2.45^{+0.24}_{-0.23}$ &$37.0^{+0.5}_{-0.5}$ & $23.5^{+0.5}_{-0.5}$\\ 
IRS 63 & 1.3 & $395.94^{+1.15}_{-1.18}$ &  $73.3^{+0.3}_{-0.3}$ & $-0.94^{+0.01}_{-0.01}$ & $4.24^{+0.07}_{-0.07}$ &$76.3^{+0.5}_{-0.5}$ & $47.8^{+0.5}_{-0.5}$\\ 
  &   3  & $58.82^{+0.54}_{-0.51}$ &  $61.9^{+1.1}_{-1.0}$ & $-1.16^{+0.01}_{-0.01}$ & $2.19^{+0.15}_{-0.13}$ &$77.3^{+2.0}_{-2.0}$ & $38.3^{+0.5}_{-0.5}$\\ 
 L1551 IRS5 N & 1.3 & $316.18^{+11.71}_{-11.28}$ &  $16.2^{+1.6}_{-2.6}$ & $-0.23^{+0.31}_{-0.21}$ & $3.68^{+1.72}_{-1.10}$ &$19.0^{+1.0}_{-0.5}$ & $13.0^{+0.5}_{-0.5}$\\  
  &   3  & $56.28^{+1.75}_{-1.66}$ &  $15.3^{+2.1}_{-3.0}$ & $-0.32^{+0.33}_{-0.23}$ & $3.05^{+1.23}_{-0.83}$ &$19.0^{+0.5}_{-0.5}$ & $12.5^{+0.5}_{-0.5}$\\ 
 L1551 IRS5 S & 1.3 & $163.99^{+3.39}_{-12.38}$ &  $12.4^{+2.0}_{-2.5}$ & $-0.07^{+0.31}_{-0.23}$ & $2.28^{+0.58}_{-0.47}$ &$19.0^{+0.5}_{-0.5}$ & $12.0^{+0.5}_{-0.5}$\\    
  &   3 & $31.21^{+0.57}_{-0.56}$ &  $11.5^{+2.0}_{-2.3}$ & $-0.19^{+0.26}_{-2.21}$ & $2.16^{+0.54}_{-0.42}$ &$18.0^{+0.5}_{-0.5}$ & $11.0^{+0.5}_{-0.5}$\\    
\hline
   \end{tabular}
   \tablefoot{
   \tablefootmark{a}{F$_{\text{tot}}$ corresponds to the total face-on flux.}
   }
\end{table*}

\begin{table*}
   \renewcommand*{\arraystretch}{1.4}
   \caption{Disk properties derived from fit to the emission using a 2D Gaussian fit}
   \label{tab:sizes_slopes_gaussian}
   \centering
   \begin{tabular}{llccccc}
   \hline\hline
   source & $\lambda$ & $F_{\nu}$ &Deconv. major FWHM & Deconv. minor FWHM&R$_{95}$ & R$_{68}$ \\ 
   & (mm) & (mJy) & (au) & (au) & (au) & (au)  \\ 

   \hline
 RCrA IRS7B b	&	1.3	&	35.64	$\pm$	0.68	&	27.54	$\pm$	0.55	&	11.7	$\pm$	0.30	&	28.65	$\pm$	0.57	&	17.66	$\pm$	0.35	\\
	&	3	&	6.32	$\pm$	0.16	&	25.37	$\pm$	0.67	&	10.4	$\pm$	0.34	&	26.40	$\pm$	0.70	&	16.27	$\pm$	0.43	\\
RCrA IRS7A	&	1.3	&	12.32	$\pm$	0.24	&	7.02	$\pm$	0.30	&	3.9	$\pm$	0.21	&	7.30	$\pm$	0.31	&	4.50	$\pm$	0.19	\\
	&	3\tablefootmark{a}			&	6.54	$\pm$	0.25	&	10.07	$\pm$	0.49	&	6.1	$\pm$	0.48	&	10.48	$\pm$	0.51	&	6.46	$\pm$	0.32	\\
CrA24	&	3	&	1.82	$\pm$	0.09	&	16.21	$\pm$	0.92	&	4.8	$\pm$	0.54	&	16.87	$\pm$	0.96	&	10.40	$\pm$	0.59	\\
CXO 34	&	1.3	&	16.27	$\pm$	0.2	&	18.53	$\pm$	0.27	&	7.7	$\pm$	0.11	&	19.28	$\pm$	0.28	&	11.88	$\pm$	0.17	\\
	&	3	&	3.62	$\pm$	0.04	&	17.60	$\pm$	0.23	&	7.6	$\pm$	0.14	&	18.31	$\pm$	0.24	&	11.28	$\pm$	0.15	\\
Elias 29	&	1.3	&	13.93	$\pm$	0.04	&	2.07	$\pm$	0.04	&	1.8	$\pm$	0.04	&	2.15	$\pm$	0.04	&	1.33	$\pm$	0.02	\\
	&	3	&	4.25	$\pm$	0.02	&	2.23	$\pm$	0.08	&	2.0	$\pm$	0.08	&	2.32	$\pm$	0.08	&	1.43	$\pm$	0.05	\\
RCrA SMM2 a	&	3	&	0.51	$\pm$	0.04	&	point source			&	point source			&	$\lesssim3$\tablefootmark{b}			&	$\lesssim2$\tablefootmark{b}			\\
RCrA SMM2 b	&	3	&	0.51	$\pm$	0.046	&	point source			&	point source			&	$\lesssim3$			&	$\lesssim2$		\\
VLA 1623 Aa	&	1.3	&	49.85	$\pm$	0.39	&	16.42	$\pm$	0.14	&	9.0	$\pm$	0.09	&	17.08	$\pm$	0.15	&	10.53	$\pm$	0.09	\\
	&	3	&	10.00	$\pm$	0.06	&	14.32	$\pm$	0.10	&	7.4	$\pm$	0.09	&	14.90	$\pm$	0.11	&	9.18	$\pm$	0.07	\\
VLA 1623 Ab	&	1.3	&	45.61	$\pm$	0.38	&	14.02	$\pm$	0.10	&	7.2	$\pm$	0.09	&	14.59	$\pm$	0.10	&	8.99	$\pm$	0.06	\\
	&	3	&	9.78	$\pm$	0.06	&	15.84	$\pm$	0.14	&	8.6	$\pm$	0.10	&	16.48	$\pm$	0.15	&	10.16	$\pm$	0.09	\\
IRAS 16293 A1	&	1.3	&		$5.90\pm0.5$		&		...	&		...		&	$\lesssim6.5$\tablefootmark{c}				&	$\lesssim4$\tablefootmark{c}			0.00	\\
	&	3	&	4.91	$\pm$	3.6	&	4.79	$\pm$	2.68	&	2.8	$\pm$	2.26	&	4.99	$\pm$	2.79	&	3.07	$\pm$	1.72	\\
IRAS 16293 A2	&	1.3	&	$6.00\pm0.6$ 	&			...	&	...		&	$\lesssim12.6$\tablefootmark{c}				&	$\lesssim7.8$\tablefootmark{c}			0.00	\\
	&	3	&	4.20	$\pm$	0.1	&	10.58	$\pm$	2.12	&	6.8	$\pm$	3.53	&	11.00	$\pm$	2.20	&	6.78	$\pm$	1.36	\\

    \hline
   \end{tabular}
   \tablefoot{
\tablefootmark{a} The 3 mm observations likely contain significant non-thermal emission, thus the disk size at this wavelength is not reliable. \\
\tablefootmark{b} The results from the fit is that the source is a point source. Upper limit corresponds to assuming the deconvolved size of the source is a third of the major beam axis. 
\tablefootmark{c} Imfit could not deconvolve the source from the beam because it was only marginally resolved in just one direction. Values in the table are thus only provided as a reference. They are calculated assuming the deconvolved FWHM of the source  is a third of the minor beam axis divided by the cosine of the inclination, the latter derived from the 3 mm observations. 
   }
\end{table*}

\section{Other predictions for the size-luminosity relation}
\label{sec:alternative_temp_profile}
To investigate further the issue of the disks being more luminous than fully optically thick passive heating predicts, we performed another test pushing the temperature to higher values by taking into account that the flaring angle $\varphi$ also changes as a function of the star luminosity and radius of the disk \citep{1997ChiangSpectral}. Following \cite{1997ChiangSpectral} to compute $\varphi$ as a function of $r$ and L$_*$, we arrive to a slightly different parametrization for the temperature such that $T=T_0(r/r_0)^{-0.43}(L_*/L_{\odot})^{0.29}$. For a 1 M$_{\odot}$\footnote{The dependece on $M_*$ on the final temperature is weak with $T\propto M^{-1/7}_*$.} and 1 L$_{\odot}$ star we get that $T_0\approx50$ K at $r_0=10$ au. With this new parametrization we get higher temperatures thus getting closer to the observed location of the Class 0 and I disks in the size-luminosity relation (Figure~\ref{fig:size_lum_varalpha}). In Figure~\ref{fig:size_lum_varalpha_lbol} we repeat this exercise but considering $L_*=L_{bol}$, instead of $L_*=L_{int}$. In both cases the discrepancy decreases but still persist for at least several sources.

\begin{figure*}[t!]
   \centering
     \includegraphics[width=0.8\textwidth]{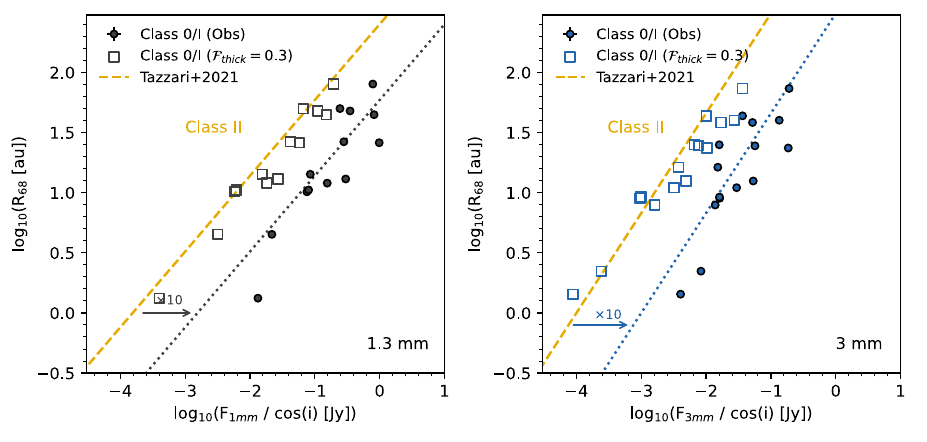}
      \caption{Comparison between the observed disk luminosity and the predicted values assuming $\mathcal{F}_{thick}=0.3$ as in \cite{2018AndrewsScalingRelations} for Class II disks. Same as Figure~\ref{fig:size_lum}. Filled circles are the observations and squares are the predicted luminosity for each disk if $\mathcal{F}_{thick}=0.3$ of the flux if optically thick. }
         \label{fig:size_lum_f0.3}
\end{figure*}

\begin{figure*}[t!]
   \centering
     \includegraphics[width=0.57\textwidth]{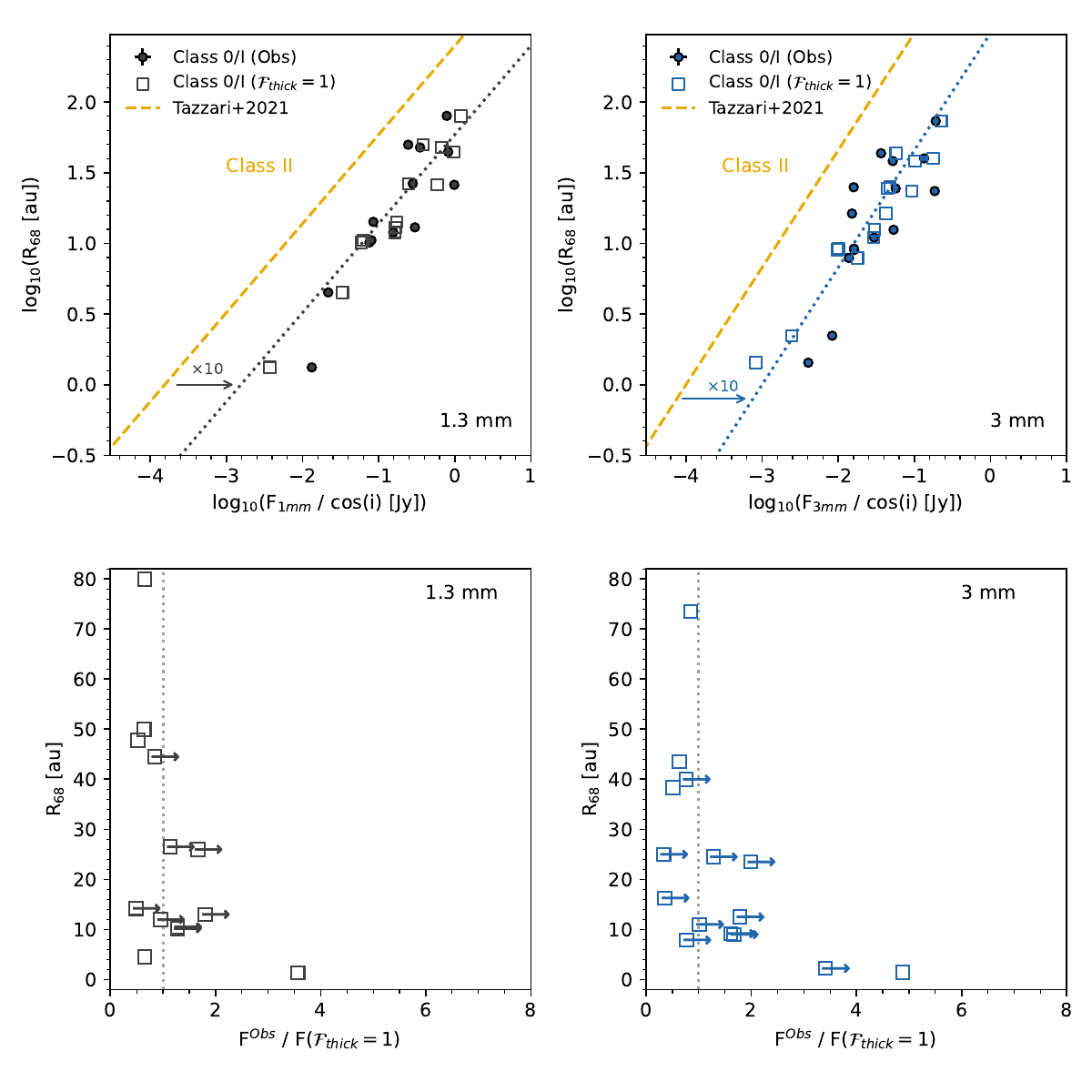}
      \caption{Comparison between the observed disks luminosity and the predicted values assuming the disks are fully optically thick, considering a non-constant flaring angle $\varphi$ and $L_*=L_{int}$. Top: Same as Figure~\ref{fig:size_lum}. Filled circles are the observations and squares are the predicted luminosity for each disk if the emission is fully optically thick. Bottom: Size versus the ratio of the observed disk luminosity to the predicted one assuming the entire disk is optically thick. Lower limits for the ratio are due to the assumed internal luminosity being an upper limit.}
         \label{fig:size_lum_varalpha}
\end{figure*}

\begin{figure*}[t!]
   \centering
     \includegraphics[width=0.57\textwidth]{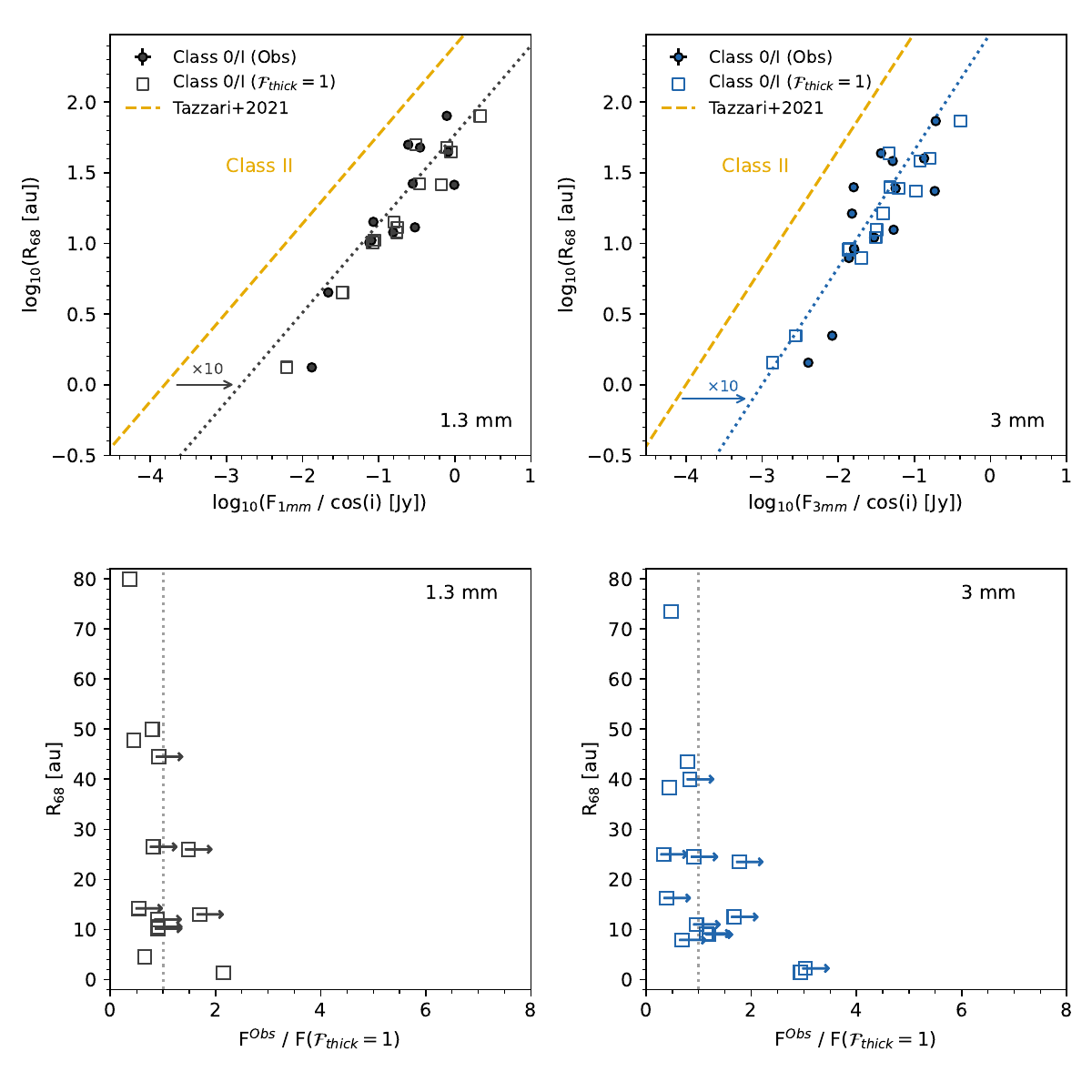}
      \caption{Comparison between the observed disks luminosity and the predicted values assuming the disks are fully optically thick, considering a non-constant flaring angle $\varphi$ and $L_*=L_{bol}$. Top: Same as Figure~\ref{fig:size_lum}. Filled circles are the observations and squares are the predicted luminosity for each disk if the emission is fully optically thick. Bottom: Size versus the ratio of the observed disk luminosity to the predicted one assuming the entire disk is optically thick. Lower limits for the ratio are due to the assumed luminosity being an upper limit.}
         \label{fig:size_lum_varalpha_lbol}
\end{figure*}

\section{Optically thick fractions at 1.3 and 3 mm}

\begin{figure*}
\centering
     \includegraphics[width=0.5\textwidth]
   {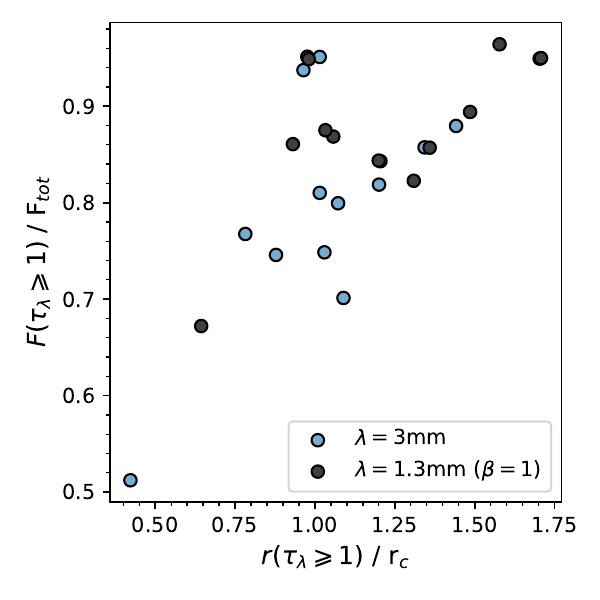}
     \caption{Fractions of the total flux and fractions of the disk characteristic radius $r_c$ containing emission with $\tau>1$ at 1.3 and 3 mm. Each symbol corresponds to an individual disk. }
     \label{fig:opt_thick_frac}
\end{figure*}

Figure~\ref{fig:opt_thick_frac} shows the fractions of the total flux and fraction of the disk radius containing emission with $\tau>1$ (left panel) at 1.3 and 3 mm.

\section{Optical depth and masses in the CBDs}
\label{sec:cb_appendix_tau}
\begin{figure*}[t!]
   \centering
     \includegraphics[width=1\textwidth]{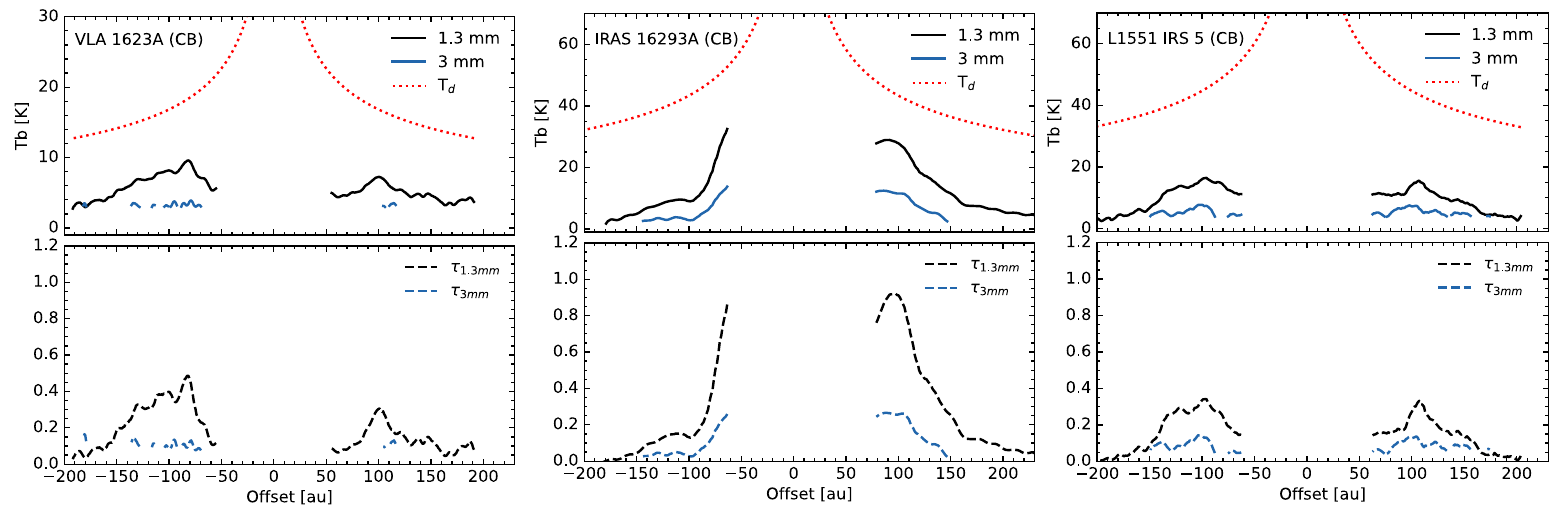}
      \caption{Intensity profiles in a cut along the major axis direction for the three CBDs in our sample observed at 1.3 and 3mm. The zero-offset corresponds to the mid-point between the circumstellar disks. Positive offsets track positive R.A. offsets. The red dotted line is the predicted temperature profile due to irradiation and the bottom panels show the resultant optical depth at each wavelength based on this temperature profile. The emission from the circumstellar disks or temperature substructures in the case of IRAS 162923A are masked out from the central region.}
         \label{fig:cb_temp_tau}
\end{figure*}

To estimate the optical depth in the circumbinary disk structures, we performed intensity cuts along the direction of the major axis of the circumbinary disk structures, centered at the midpoint between the circumstellar disks. The intensity at each location is averaged over a beam. For the P.A. we used $36^{\circ}$, $48^{\circ}$, and $165^{\circ}$, for VLA1623A, IRAS 16293A, and L1551 IRS 5, respectively. We masked out emission from the individual CBDs at the center. In the case of IRAS 16293A we also masked out the emission coming from hot dust spots features which are due to shocks and thus their temperature cannot be modeled considering only irradiation. The resultant intensity profiles are shown in Figure~\ref{fig:cb_temp_tau}. To infer the optical depth we assumed that that the intensity is given by a modified black body and assumed a temperature profile from irradiation following the parametrization discussed in discussed in Section~\ref{sec:alternative_temp_profile}. The luminosities were taken from Table~\ref{table:source_prop}. The resultant optical depths are shown in the bottom panels of Figure~\ref{fig:cb_temp_tau}. We note that using other parametrization more typical of envelopes (e.g., \citealt{2019GalametzLowdust}) results in higher temperatures and thus even lower optical depths. However, assuming a colder temperature parametrization such as that in \cite{2018AndrewsScalingRelations}, also discussed in Section~\ref{sec:disc_optdepth_size_luminosity} in this work, results in temperatures in some regions that go below the observed brightness temperatures. In order to improve the derivation of the temperature and optical depth, in a more model independent way, more wavelengths are required.

To estimate the mass in the VLA1623A and L1551 IRS5 CBDs, we measure the 1.3 mm integrated flux in the CBDs, and use the optically thin approximation:

\begin{equation}
    M=\frac{d^2S_{\nu}}{B_{\nu}(T_{d})\kappa_{\nu}}
    \label{eq:massfromdust}
\end{equation}

\noindent where $S_{\nu}$ is the integrated flux density, $B_{\nu}$ is the Planck function, $T_d$ the dust temperature, and $d$ is the distance. Given the low optical depths for these CBDs  ($\tau_{1.3mm}\lesssim0.5$ Appendix~\ref{sec:cb_appendix_tau}), the mass underestimations due to optically thin assumptions are less than 25\%, which is lower than the typical uncertainties due to the choice of dust opacity.  For the temperature, we calculated the mean value along the major axis weighted by the 1.3 mm intensity (see Figure~\ref{fig:cb_temp_tau}). The obtained values are $T_d=17$ K and $T_d=44$ K for VLA1623 and L1551 IRS5, respectively. To be consistent with the CSD mass measurements in this work, we use the same value for $\kappa_{3mm}=1$ cm$^{2}$ g$^{-1}$ and extrapolate this value to 1.3 mm (223 GHz) according to the mean $\beta$ values measured for these CBDs  (Section~\ref{sec:cb_beta}). These are $\beta=1.1$ for L1551 IRS5 and $\beta=1.37$ for VLA 1623. For IRAS 16293 we used the mass values derived in \cite{2022Maureirahotspots} and corrected for the opacity as above using $\beta=1.5$. This yielded values that are 3.7$\times$ smaller than those in \cite{2022Maureirahotspots}.

\end{appendix}

\end{document}